\documentclass{article}

\PassOptionsToPackage{numbers, compress}{natbib}
\usepackage[preprint]{neurips_2026}

\usepackage[utf8]{inputenc} 
\usepackage[T1]{fontenc}    
\usepackage{hyperref}       
\usepackage{url}            
\usepackage{booktabs}       
\usepackage{amsfonts}       
\usepackage{nicefrac}       
\usepackage{microtype}      
\usepackage{xcolor}         
\usepackage{amsmath}
\usepackage{enumitem}
\usepackage{multirow}
\newcommand{\cmark}{\ding{51}} 
\newcommand{\xmark}{\ding{55}} 
\usepackage{amssymb}
\usepackage{array}
\usepackage{graphicx} 
\usepackage{color}
\usepackage{makecell} 
\usepackage{float}
\usepackage{subcaption}
\usepackage{colortbl}
\usepackage{placeins}
\usepackage{comment}
\usepackage{pifont}
\usepackage{makecell}
\definecolor{visionbg}{HTML}{E8F0FE}
\definecolor{audiobg}{HTML}{FFF3E0}
\definecolor{unifiedbg}{HTML}{E8F5E9}

\title{FLARE: Full-Modality Long-Video Audiovisual Retrieval Benchmark with User-Simulated Queries}

\author{%
\begin{tabular}{@{}ccccc@{}}
Qijie You\textsuperscript{1}\thanks{Equal contribution.} &
Hao Liang\textsuperscript{2,4}\footnotemark[1]\hspace{0.5em}\thanks{Project leader.} &
Mingrui Chen\textsuperscript{3} &
Bohan Zeng\textsuperscript{2} &
Meiyi Qiang\textsuperscript{2}
\\[0.4em]
\multicolumn{5}{c}{%
  \begin{tabular}{@{}cc@{}}
  Zhenhao Wong\textsuperscript{2} &
  \hspace{3em}Wentao Zhang\textsuperscript{2,4}\thanks{Corresponding author.}
  \end{tabular}
}
\\[0.6em]
\multicolumn{5}{c}{\normalfont\textsuperscript{1}University of Science and Technology Beijing}
\\[-0.1em]
\multicolumn{5}{c}{\normalfont\textsuperscript{2}Peking University}
\\[-0.1em]
\multicolumn{5}{c}{\normalfont\textsuperscript{3}Institute of Automation, Chinese Academy of Sciences}
\\[-0.1em]
\multicolumn{5}{c}{\normalfont\textsuperscript{4}Zhongguancun Academy}
\end{tabular}
}

\begin{document}

\maketitle


\title{FLARE: Full-Modality Long-Video Audiovisual Retrieval Benchmark \\ with User-Simulated Queries}

\begin{abstract}
As video becomes increasingly central to information dissemination and multimodal large language models (MLLMs) continue to advance, evaluating video retrieval has become increasingly important. In realistic search scenarios, this requires matching short user queries to long-form content using both visual and auditory evidence. Yet existing retrieval benchmarks are still dominated by short clips, single modalities, and caption-based evaluation. We introduce FLARE, a full-modality long-video audiovisual retrieval benchmark with user-simulated queries. Built from 399 carefully screened Video-MME videos (10--60\,min, 225.4\,h) to ensure source quality and diversity, FLARE contains 87,697 clips annotated with vision, audio, and unified audiovisual captions, together with 274,933 user-style queries. Cross-modal queries are further filtered by a hard bimodal constraint, requiring retrieval to fail under either modality alone but succeed when both are combined. FLARE evaluates models under two regimes, caption-based and query-based retrieval, across vision, audio, and unified audiovisual settings. Experiments with 15 representative retrievers show that user-style queries substantially change model behavior, strong caption-based performance does not always transfer to query-based retrieval, and audio--language alignment remains a key bottleneck for unified audiovisual retrieval. Our code and data are released at \url{https://flarebench.github.io/}
\end{abstract}


\section{Introduction}
\label{sec:introduction}

Video has become one of the dominant media for information dissemination on the modern Internet, and building high-quality video retrieval systems is therefore an increasingly important task. The rapid progress of multimodal large language models~\citep{qwen3vl,qwen3omni} has further elevated this need: video retrieval is now considered a key capability of MLLMs, since it directly reflects how well these models align long visual and auditory streams with natural-language intent. In a realistic search scenario, queries are fundamentally \emph{audiovisual} (depending on both modalities), \emph{long-context} (the target lies within a video spanning tens of minutes), and \emph{user-shaped} (much shorter and more partial than a full description). However, public benchmarks still fall short of these properties: they focus on short clips, evaluate one modality at a time, and measure models against detailed captions rather than realistic queries.

Most existing video--text and audio--text benchmarks were built around independently collected short clips, with later work mainly scaling up data. LoVR~\citep{cai2025lovr} took a step toward long videos but remains single-modal, while audiovisual benchmarks instead target QA~\citep{videomme,jointavbench}, classification~\citep{vggsound}, or composed retrieval~\citep{omnicvr}---none of which probes whether unified multimodal models can actually \emph{retrieve} long videos from realistic queries. Equally important, all of these benchmarks evaluate models against the same form of text used during construction, leaving it unclear whether reported gains transfer to the shorter, partial queries that real users issue.

To fill this gap, we introduce \textbf{FLARE}, the first \emph{F}ull-modality \emph{L}ong-video \emph{A}udiovisual \emph{R}etrieval b\emph{E}nchmark with user-simulated queries. FLARE comprises 399 long-form videos (10--60\,min, 225.4\,h) from Video-MME~\citep{videomme}, segmented into 87,697 clips, each annotated with vision, audio, and unified audiovisual captions, plus 274,933 user-style queries across three modality types. Every cross-modal query satisfies a \emph{hard bimodal constraint} (Eq.~\ref{eq:hard_bimodal}): it must \emph{fail} unimodal retrieval and only \emph{succeed} when both modalities are combined, ensuring genuine audiovisual reasoning. Built on top of these annotations, FLARE further introduces a \emph{dual-regime} evaluation protocol that benchmarks every model under both a caption-based regime (detailed captions) and a query-based regime (user-style queries) on the same gallery, isolating the impact of query formulation on model assessment.

Benchmarking 15 representative contrastive and LLM-based retrievers~\citep{radford2021clip,siglip2,metaclip2,videoclipxl,qwen3vl_emb,msclap,clap_laion,m2dclap,glap,aurola,girdhar2023imagebind,zhu2024languagebind,perception_meta,wave} reveals two phenomena that demonstrate the value of FLARE. First, switching from captions to user-style queries causes the previously top-performing model in \emph{each} of the three modalities to suffer a drop of around 20--40\% in Recall@1, showing that user queries, although derived from the same content, pose a much greater challenge than information-rich captions and thus constitute an indispensable complement to caption-based evaluation. Second, comparing single-modal and full-modality settings reveals that audio retrieval is consistently the bottleneck of unified audiovisual models; strengthening audio--language alignment and designing better cross-modal fusion mechanisms thus remain important open challenges that single-modality benchmarks fundamentally cannot expose.

Our contributions are summarized as follows:
\begin{itemize}[leftmargin=*, itemsep=2pt]
    \item We present FLARE, which, to the best of our knowledge, is the first full-modality long-video retrieval benchmark. A highly automated generation pipeline, complemented by rigorous human-assisted review, yields a high-quality benchmark of 399 long-form videos and 87,697 clips with vision, audio, and unified captions that jointly supports vision, audio, and unified audiovisual retrieval.
    \item We further construct 274,933 user-style queries across three modality types, making FLARE, to the best of our knowledge, the first benchmark to introduce user-simulated queries with a hard bimodal constraint. Each query undergoes a rigorous validation process, ensuring that single-modality queries remain strongly relevant to the original captions while reflecting user-like search behavior, and that cross-modal queries further require information from both modalities.
    \item We comprehensively evaluate 15 baselines and reveal differences and deficiencies at two levels---\emph{caption vs.\ query} and \emph{single-modal vs.\ unified audiovisual}---highlighting the limitations of current research.
\end{itemize}


\section{Related Work}
\label{sec:related_work}

\subsection{Single-Modality Retrieval Benchmarks}
\label{sec:rw_single_modality}

Earlier video--text and audio--text retrieval benchmarks, such as MSR-VTT~\citep{xu2016msrvtt}, MSVD~\citep{chen2011collecting}, and VATEX~\citep{wang2019vatex} on the visual side, and AudioCaps~\citep{audiocaps}, Clotho~\citep{drossos2019clotho}, MusicCaps~\citep{musiccaps}, and WavCaps~\citep{mei2024wavcaps} on the audio side, were largely built on independent short clips, with subsequent improvements mainly focusing on data scale and diversity. Such settings lack long-video context as a source of difficulty and are not well aligned with realistic search scenarios. With the rapid development of multimodal LLMs, recent work represented by LoVR~\citep{cai2025lovr} extends visual retrieval to long videos and introduces a more challenging temporal context, but it still operates within a single modality. As unified multimodal models become a central trend, building a video--text retrieval benchmark that simultaneously supports single-modality and unified-modality evaluation has become increasingly important.

\subsection{Joint Audiovisual Benchmarks}
\label{sec:rw_joint_audiovisual}

As unified multimodal models continue to advance, evaluating a model's joint audiovisual capability has become a central focus of multimodal research. Some existing cross-modal benchmarks, such as Video-MME~\citep{videomme} and JointAVBench~\citep{jointavbench}, concentrate on QA-style evaluation of joint audio--visual reasoning, evaluating MLLMs via multiple-choice questions rather than embedding-based retrieval. Others, such as VGGSound~\citep{vggsound} and OmniCVR~\citep{omnicvr}, focus on audio classification, where vision is treated more as auxiliary information, or on composed retrieval tasks where the model retrieves a target video given a source video and a textual modification. In contrast, we construct the first long-video retrieval benchmark that jointly covers vision, audio, and unified audiovisual settings, and we further build a user-style query dimension on top of detailed captions, which significantly increases the realism and difficulty of FLARE. Table~\ref{tab:benchmark_comparison} summarizes the detailed comparison.

\begin{table}[t]
  \centering
  \caption{Comparison with existing benchmarks. \textbf{Construction}: H = Human, A = Automated, H+A = Human\,+\,Automated.}
  \label{tab:benchmark_comparison}
  \footnotesize
  \setlength{\tabcolsep}{2pt}
  \begin{tabular}{ll ccccc}
  \toprule
  \textbf{Category} & \textbf{Benchmark} & \textbf{Units} & \textbf{Duration} & \textbf{Construction} & \textbf{Format} & \textbf{Query} \\
  \midrule
  \multirow{4}{*}{\textbf{Vision}}
    & MSR-VTT\,\citep{xu2016msrvtt}               & 10K   & $\sim$15\,s   & H   & Caption  & \textcolor{red!70!black}{\xmark} \\
    & MSVD\,\citep{chen2011collecting}              & 1,970   & $\sim$10\,s   & H   & Caption  & \textcolor{red!70!black}{\xmark} \\
    & VATEX\,\citep{wang2019vatex}                  & 41K & $\sim$10\,s  & H   & Caption  & \textcolor{red!70!black}{\xmark} \\
    & LoVR\,\citep{cai2025lovr}                        & 40,804   & $\sim$18\,s & H+A & Caption  & \textcolor{red!70!black}{\xmark} \\
  \midrule
  \multirow{4}{*}{\textbf{Audio}}
    & AudioCaps\,\citep{audiocaps}             & 51K    & $\sim$10\,s   & H   & Caption  & \textcolor{red!70!black}{\xmark} \\
    & Clotho\,\citep{drossos2019clotho}         & 6,974     & 15--30\,s     & H   & Caption  & \textcolor{red!70!black}{\xmark} \\
    & MusicCaps\,\citep{musiccaps}              & 5,521   & 10\,s         & H   & Caption  & \textcolor{red!70!black}{\xmark} \\
    & WavCaps\,\citep{mei2024wavcaps}           & 403K   & $\sim$67.6\,s        & A   & Caption  & \textcolor{red!70!black}{\xmark} \\
  \midrule
  \multirow{6}{*}{\textbf{Vision+Audio}}
    & VGGSound\,\citep{vggsound}                & 200K   & 10\,s         & A   & Caption & \textcolor{red!70!black}{\xmark} \\
    & Video-MME\,\citep{videomme}               & 900   & 11\,s--1\,h   & H   & QA    & \textcolor{red!70!black}{\xmark} \\
    & OmniCVR\,\citep{omnicvr}                  & 160K     & $\sim$11.8\,s            & H+A & Caption & \textcolor{red!70!black}{\xmark} \\
    & JointAVBench\,\citep{jointavbench}        & 1,046   & $\sim$97\,s            & H+A & QA    & \textcolor{red!70!black}{\xmark} \\
  \cmidrule(lr){2-7}
    & \textbf{Ours (FLARE-Clip)}                     & \textbf{87,697} clips  & $\sim$\textbf{9.2\,s} & H+A & Caption+Query & \multirow{2}{*}{\textcolor{green!60!black}{\cmark}} \\
    & \textbf{Ours (FLARE-Video)}                    & \textbf{399} videos  & $\sim$\textbf{33.9\,min} & H+A & Caption &  \\
  \bottomrule
  \end{tabular}
  
  \vspace{2pt}
  \end{table}


\section{Dataset Construction}
\label{sec:dataset_construction}

This section describes how we construct high-quality FLARE data across three modality settings---vision, audio, and unified audiovisual---and two textual forms: detailed captions and user-simulated queries. We use a highly automated pipeline supplemented by targeted human review to ensure data quality while maintaining scalability (Appendix~\ref{app:human_annotation}); the main prompts used in the automated stages are reported in Appendix~\ref{app:prompts}.

\begin{figure}[!t]
    \centering
    \includegraphics[width=1\linewidth]{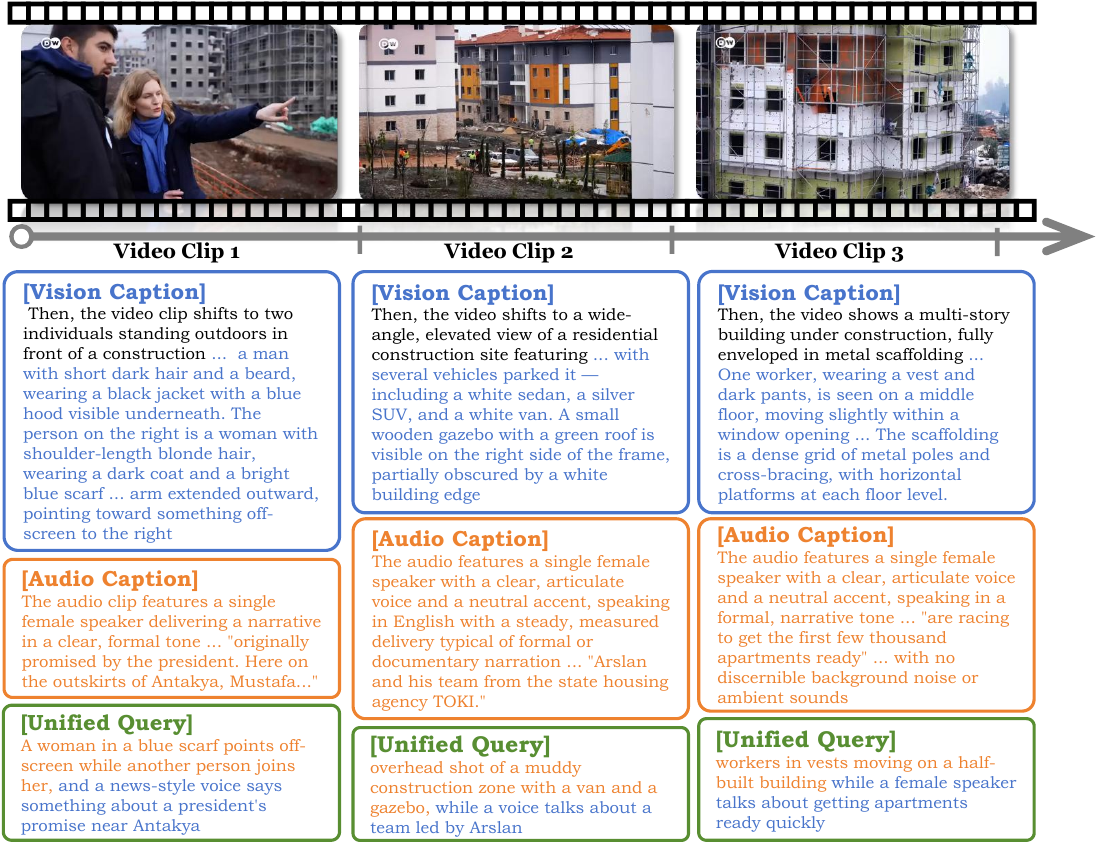}
    \caption{Data Case. Due to space constraints, many parts are omitted.}
    \label{fig:data_case}
\end{figure}

\subsection{Video Source and Clip Extraction}
\label{sec:clip_extraction}

\paragraph{Video source selection.}
We build upon Video-MME~\citep{videomme}, which covers 6 visual domains and 30 subfields with videos ranging from 11\,s to 1\,h, incorporating rich multimodal signals (visual content, subtitles, and audio tracks). To focus on challenging long-form retrieval, we retain only videos exceeding 10 minutes.

\paragraph{Visual scene segmentation.}
Each video $v$ is decomposed into temporally contiguous clips $v = \{c_1, c_2, \ldots, c_N\}$ via the \texttt{ContentDetector} of PySceneDetect~\citep{pyscenedetect}. For frames $\{f_1, \ldots, f_M\}$, a content change score $\delta_i$ is computed between consecutive frames $f_i$ and $f_{i+1}$; a scene boundary is placed when $\delta_i > \tau$, subject to a minimum duration $d_{\min}$. We set $\tau = 30.0$ and $d_{\min} = 3\,\text{s}$ based on parameter sweeps with manual inspection.

\paragraph{Modality triage and audio-driven segmentation.}
Clips that remain longer than one minute after visual segmentation typically derive their semantic continuity from the audio track (e.g., lectures, narration) rather than visual change, motivating segmentation from the audio side. To verify this premise, we score each such clip with Qwen3-VL-235B-A22B-Instruct~\citep{qwen3vl}, which assigns an audio importance score $s_{\text{aud}} \in [0, 10]$; clips with $s_{\text{aud}} \geq 6$ are routed to audio-driven segmentation, while borderline cases (Appendix~\ref{app:modality_verification}) are adjudicated by human annotators.

\textit{Audio semantic segmentation.}\quad We transcribe each clip with Qwen3-ASR-1.7B~\citep{qwen3asr} to obtain word-level timestamps, and use Qwen3-235B-A22B-Instruct~\citep{qwen3_235b} to partition the transcript into topically coherent paragraphs $\{S_1, \ldots, S_P\}$, each mapped back to a temporal span $[t^s_p, t^e_p]$. Spans shorter than $d_{\min}^{\text{sem}} = 1\,\text{s}$ are merged with neighbors.

\textit{Spectral novelty detection.}\quad The LLM-based splitter above relies on speech transcripts and therefore cannot place boundaries at non-speech acoustic events (e.g., music shifts, applause, ambient changes). As a complementary signal, we run spectral novelty detection over the mel-spectrogram $\mathbf{M}$ and combine two novelty functions:

\noindent\textit{(i) Spectral flux:}
\begin{equation}
\label{eq:flux}
\Phi(t) = \sum_{b=1}^{B} \max\!\bigl(\log M_{b,t} - \log M_{b,t-1},\; 0\bigr).
\end{equation}

\noindent\textit{(ii) KL divergence}, measuring the per-frame spectral-shape shift after normalizing each frame to $\mathbf{p}_t = \mathbf{M}_{:,t} / \|\mathbf{M}_{:,t}\|_1$:
\begin{equation}
\label{eq:kl}
D_{\mathrm{KL}}(t) = \sum_{b=1}^{B} p_{b,t}\, \log \frac{p_{b,t}}{p_{b,t-1}}.
\end{equation}

Both curves are standardized with a robust (MAD-based) z-score and fused with equal weight:
\begin{equation}
\label{eq:novelty}
\mathcal{N}(t) = \tfrac{1}{2}\,\hat{\Phi}(t) + \tfrac{1}{2}\,\hat{D}_{\mathrm{KL}}(t).
\end{equation}
Local peaks of $\mathcal{N}$ exceeding $z_{\min} = 5$ with a minimum inter-peak gap of $1\,\text{s}$ form candidate boundaries; resulting segments shorter than $d_{\min}^{\text{spec}} = 3\,\text{s}$ are merged with neighbors.

\paragraph{Final human review.}
To guarantee that every retained clip is retrieval-friendly---short enough to carry a single coherent topic and exhibiting non-trivial audiovisual variation---clips still exceeding 2 minutes after automated segmentation are routed to human review, where annotators manually split them along the dominant modality or discard clips lacking meaningful audio-visual content (full protocol in Appendix~\ref{app:clip_verification}).

\subsection{Multimodal Caption Generation}
\label{sec:caption_generation}

Each clip is annotated with a visual caption, an audio caption, and a unified multimodal caption. We further aggregate clip-level captions into video-level descriptions through hierarchical merging, with multi-layer quality assurance throughout.

\paragraph{Visual and audio captioning.}
We generate modality-specific captions with strong multimodal models: Qwen3-VL-235B-A22B-Instruct~\citep{qwen3vl} for visual content and Qwen3-Omni-30B-A3B-Instruct~\citep{qwen3omni} for audio content. Visual captions describe visible objects, actions, attributes, spatial relations, and other visual cues, while audio captions capture speech, speakers, music, and other auditory events; both faithfully and comprehensively reflect the corresponding modality in the video.

\paragraph{Caption quality assurance.}
We filter visual and audio captions with EVQAScore~\citep{evqascore} and BRACEScore~\citep{brace}, respectively, supplemented by semantic coherence checks using Qwen3-235B-A22B-Instruct. The quality thresholds are calibrated on a small human-reviewed pilot set, and captions falling below the thresholds or flagged by the LLM are manually reviewed and corrected (Appendix~\ref{app:caption_qa}).

\paragraph{Unified caption generation.}
Given the quality-assured vision and audio captions, Qwen3-Omni-30B-A3B-Instruct~\citep{qwen3omni} generates a high-quality unified caption that preserves information from both modalities and naturally integrates it according to the video content.

\paragraph{Hierarchical caption merging.}
\label{sec:caption_merging}
Directly captioning entire long videos leads to incomplete descriptions. We instead construct video-level captions bottom-up: clip-level unified captions are sorted by scene order and partitioned into clusters of $k{=}10$. Within each cluster, an LLM smooths only the \emph{adjacent paragraph boundaries}. Let $p_{c_i}^{\text{tail}}$ and $p_{c_{i+1}}^{\text{head}}$ denote the last and first paragraphs of consecutive captions. The merged result is:
\begin{equation}
\label{eq:merge}
\mathrm{cap}_{\text{merged}} = p_{c_i}^{\text{rest}} + \mathrm{LLM}\!\left(p_{c_i}^{\text{tail}} \,\|\, p_{c_{i+1}}^{\text{head}}\right) + p_{c_{i+1}}^{\text{rest}}
\end{equation}
where $\mathrm{LLM}(\cdot)$ rewrites the transition using temporal connectives while preserving all factual details. Cluster-level captions are then concatenated to form the final video-level caption $\mathrm{cap}_v$.

\begin{figure}[!t]
    \centering
    \includegraphics[width=1\linewidth]{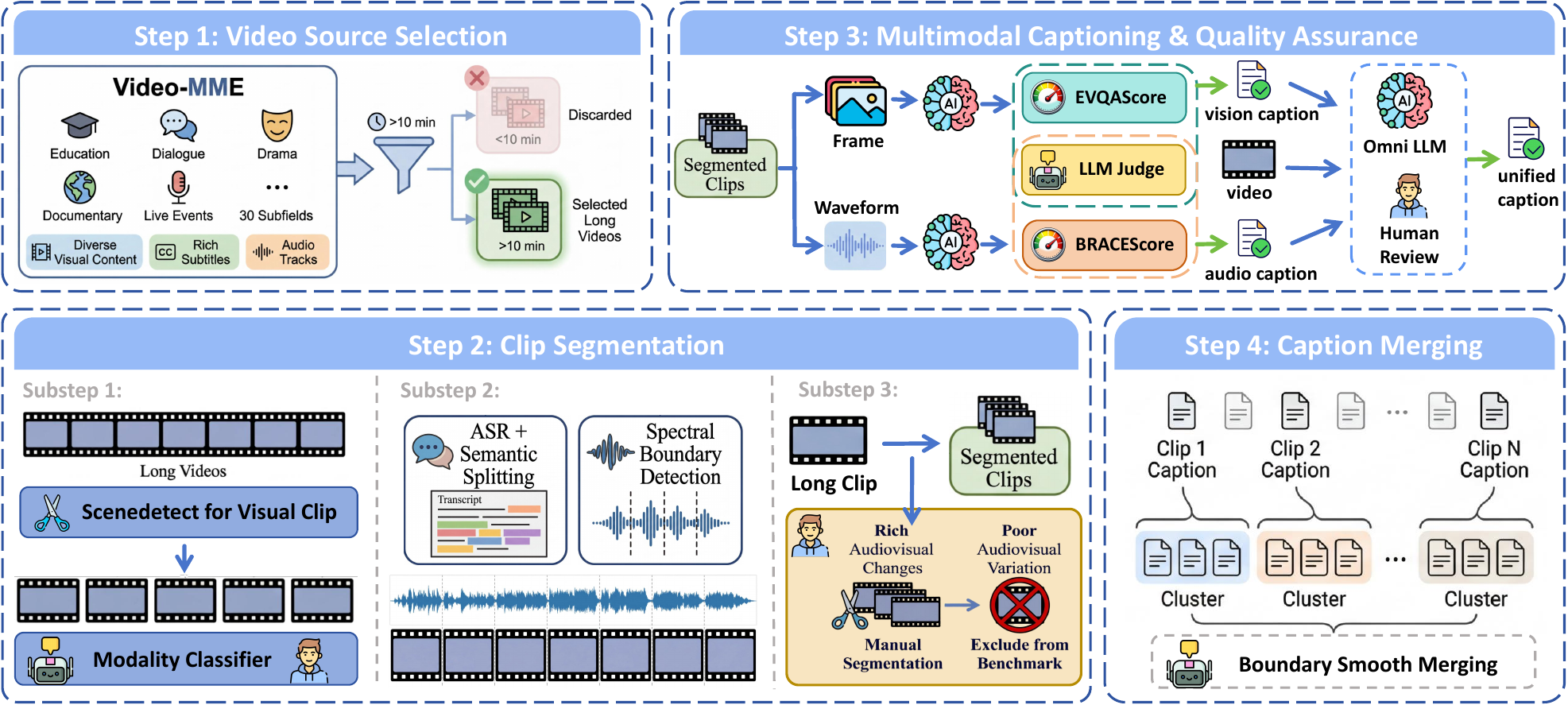}
    \caption{Caption Construct}
    \label{fig:caption_construct}
\end{figure}

\begin{figure}[!t]
    \centering
    \includegraphics[width=0.9\linewidth]{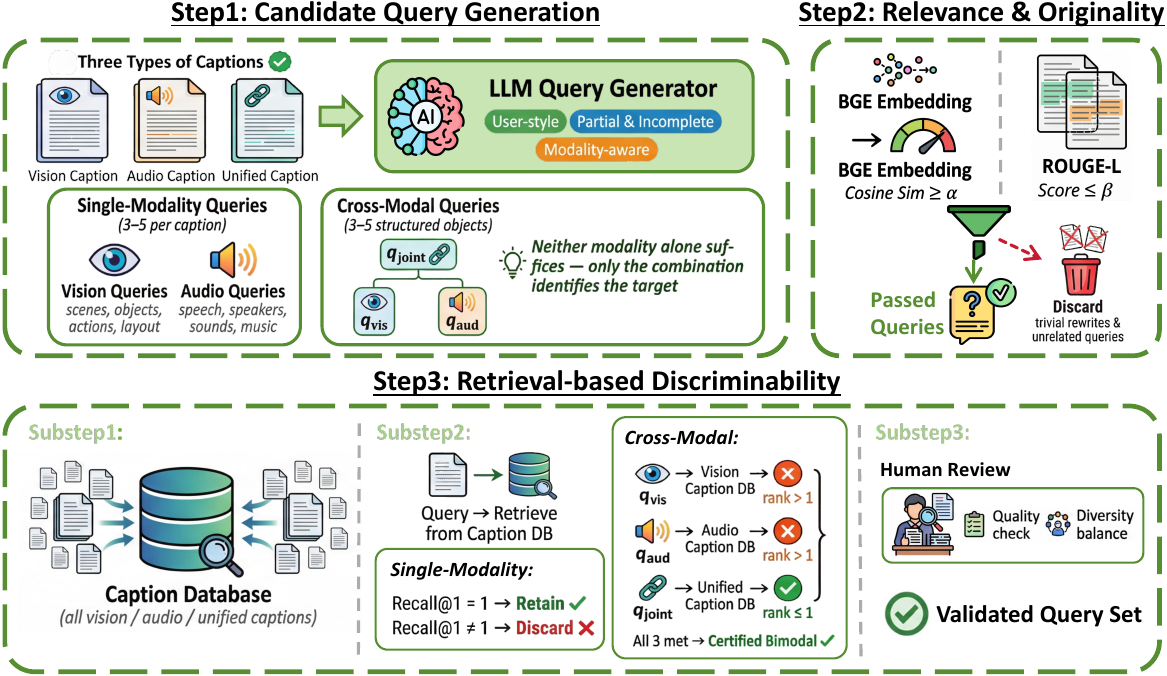}
    \caption{Query Construct}
    \label{fig:query_construct}
\end{figure}

\subsection{Multimodal Query Generation}
\label{sec:query_generation}

In real retrieval scenarios, users often do not provide complete or exhaustive descriptions. To make FLARE closer to realistic search behavior, we construct a high-quality dataset of simulated user queries through the following steps.

\paragraph{Candidate generation.}
For vision-only and audio-only queries, Qwen3-235B-A22B-Instruct rewrites the corresponding modality caption into 3--5 concise search queries that preserve the target semantics while omitting non-essential details. For cross-modal queries, the same model produces a joint query $q_{\text{joint}}$ together with its visual component $q_{\text{vis}}$ and audio component $q_{\text{aud}}$, enabling us to explicitly test whether both modalities are needed.

\paragraph{Relevance and non-copy filtering.}
A useful query should remain semantically tied to its source caption without simply paraphrasing it. We therefore compare each candidate query $q$ with its source caption $c$ using semantic similarity from BGE-Multilingual-Gemma2~\citep{bge_m3} and lexical overlap measured by ROUGE-L~\citep{lin2004rouge}, retaining only candidates that satisfy:
\begin{equation}
\label{eq:stage1}
\mathrm{sim}(q, c) \geq \theta_{\mathrm{sim}} \quad \text{and} \quad \mathrm{R}_L(q, c) \leq \theta_{\mathrm{rouge}},
\end{equation}
where $\theta_{\mathrm{sim}} = 0.4$ enforces semantic relevance and $\theta_{\mathrm{rouge}} = 0.2$ discourages caption copying.

\paragraph{Retrieval-based validation.}
We further require each query to retrieve its source clip from the full benchmark gallery. Single-modality queries must rank the target clip first under the corresponding modality-specific captions. Cross-modal queries are held to a stricter \textbf{hard bimodal constraint}: let $\mathrm{rank}_v$, $\mathrm{rank}_a$, and $\mathrm{rank}_j$ denote the target rank under the vision, audio, and unified caption spaces. We retain a cross-modal query only if
\begin{equation}
\label{eq:hard_bimodal}
\underbrace{\mathrm{rank}_v > K_v}_{\text{vision-only fails}} \;\wedge\; \underbrace{\mathrm{rank}_a > K_a}_{\text{audio-only fails}} \;\wedge\; \underbrace{\mathrm{rank}_j \leq K_j}_{\text{joint succeeds}},
\end{equation}
with $K_v = K_a = K_j = 1$. Thus, every retained cross-modal query is insufficient under either modality alone but succeeds when audiovisual information is combined. A held-out human review further verifies the naturalness and correctness of the curated queries (Appendix~\ref{app:final_quality}).

\subsection{Dataset Statistics}
\label{sec:statistics}

\begin{figure}[t]
\centering
\includegraphics[width=\textwidth]{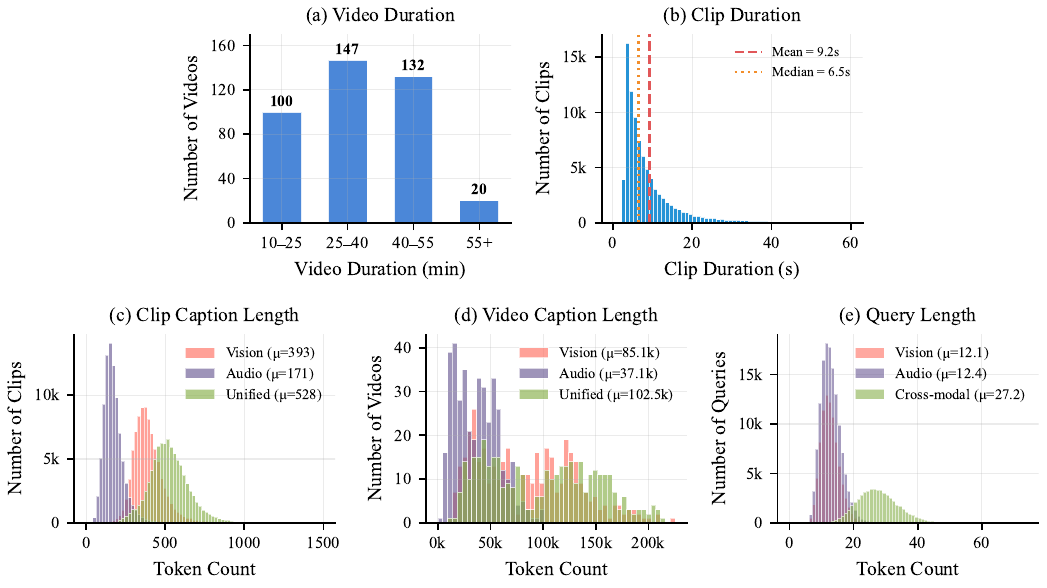}
\caption{Dataset statistics distributions. (a)~Video duration grouped by range. (b)~Clip duration in seconds. (c)~Clip-level caption token length for vision, audio, and unified modalities. (d)~Video-level caption token length across three modality types. (e)~Query token length distribution across vision, audio, and cross-modal query types.}
\label{fig:statistics}
\end{figure}

Figure~\ref{fig:statistics} summarizes the main statistics of FLARE. The video-duration distribution confirms that the benchmark is built from genuinely long-form content: all 399 source videos exceed 10 minutes, totaling 225.4 hours with a mean duration of 33.9 minutes. This long-video setting makes retrieval substantially harder than short-clip benchmarks, since models must distinguish targets within extended temporal context. At the same time, the clip-duration distribution shows that our segmentation is fine-grained (mean 9.2\,s, median 6.5\,s), producing 87,697 retrieval units that are short enough to preserve coherent semantics while still yielding a large and challenging gallery. On top of these clips, our curation pipeline yields 274,933 user-style queries, including 86,350 vision-only, 135,003 audio-only, and 53,580 cross-modal queries; the cross-modal subset is smaller because of the strict hard bimodal constraint (Eq.~\ref{eq:hard_bimodal}).

The length distributions further characterize the annotation design. Both clip-level and video-level captions are relatively long, indicating that they contain detailed descriptions rather than sparse labels. Unified captions are longer than vision-only or audio-only captions at both levels, which reflects their role in combining information from both modalities. In contrast, user-style queries are much shorter than captions, matching realistic search behavior and creating an additional challenge: models must retrieve the correct clip from partial, underspecified cues rather than from exhaustive descriptions. Cross-modal queries are also longer than single-modal queries, consistent with their need to encode complementary visual and auditory evidence.

\definecolor{visionbg}{HTML}{E8F0FE}
\definecolor{audiobg}{HTML}{FFF3E0}
\definecolor{unifiedbg}{HTML}{E8F5E9}

\section{Experiments}
\label{sec:experiments}

\subsection{Evaluation Protocol}
\label{sec:exp_setup}

We report Recall@1/5/10 computed via cosine similarity. The benchmark spans two orthogonal axes: \emph{modality scope}---vision, audio, and vision{+}audio---and \emph{query regime}---\emph{caption-based} (using the detailed captions of Section~\ref{sec:caption_generation}) and \emph{query-based} (using the curated user-style queries of Section~\ref{sec:query_generation}). Each configuration is evaluated on four directions: Text$\to$Clip (T$\to$C), Text$\to$Video (T$\to$V), Clip$\to$Text (C$\to$T), and Video$\to$Text (V$\to$T); the query regime is restricted to clip-level tasks since queries are generated per clip. For every model we use the strongest publicly available checkpoint and follow the official codebase and default configuration. We benchmark 15 models (see Table~\ref{tab:baselines}) covering vision (CLIP~\citep{radford2021clip}, SigLIP2~\citep{siglip2}, MetaCLIP-2~\citep{metaclip2}, VideoCLIP-XL-v2~\citep{videoclipxl}, Qwen3-VL-Emb-8B~\citep{qwen3vl_emb}), audio (MS-CLAP~\citep{msclap}, LAION-CLAP~\citep{clap_laion}, M2D-CLAP~\citep{m2dclap}, GLAP~\citep{glap}, Aurola-7B~\citep{aurola}), and vision{+}audio (ImageBind~\citep{girdhar2023imagebind}, LanguageBind~\citep{zhu2024languagebind}, Perception AV Large~\citep{perception_meta}, Wave-7B~\citep{wave}).

\subsection{Caption-Based Retrieval Results}
\label{sec:caption_results}

Table~\ref{tab:caption_results} presents caption-based retrieval results. Two findings stand out. First, \textbf{LLM-based embedding models consistently outperform contrastive baselines}: Qwen3-VL-Emb-8B reaches 80.27\% T$\to$C R@1 in the vision setting, Aurola-7B reaches 73.02\% in the audio setting, and Wave-7B reaches 65.51\% in the unified audiovisual setting, substantially exceeding the strongest contrastive model in each category. Second, \textbf{audio--language retrieval remains far behind vision--language retrieval under contrastive training}. The best contrastive audio model, M2D-CLAP, obtains only 0.56\% T$\to$C R@1, far below CLIP ViT-B/32 (7.98\%) and VideoCLIP-XL-v2 (47.28\%). This large gap suggests that current audio--language representations are still much less retrieval-ready than their vision--language counterparts. Video-level retrieval is easier because the candidate pool is smaller, with LLM-scale models approaching saturation, yet it remains challenging for smaller contrastive baselines because it requires aligning long media sequences with long textual descriptions.

\begin{table}[!t]
\centering
\caption{Caption-based retrieval results (Recall@$K$, \%). Models are grouped by modality; \textbf{bold} indicates best within each group.}
\label{tab:caption_results}
{%
\scriptsize
\renewcommand{\arraystretch}{0.9}
\setlength{\tabcolsep}{2pt}
\resizebox{\textwidth}{!}{%
\begin{tabular}{l ccc ccc ccc ccc}
\toprule
 & \multicolumn{3}{c}{Text$\to$Clip} & \multicolumn{3}{c}{Text$\to$Video} & \multicolumn{3}{c}{Clip$\to$Text} & \multicolumn{3}{c}{Video$\to$Text} \\
\cmidrule(lr){2-4} \cmidrule(lr){5-7} \cmidrule(lr){8-10} \cmidrule(lr){11-13}
Model & R@1 & R@5 & R@10 & R@1 & R@5 & R@10 & R@1 & R@5 & R@10 & R@1 & R@5 & R@10 \\
\midrule
\rowcolor{visionbg} \multicolumn{13}{c}{\textbf{Vision}} \\
CLIP ViT-B/32          &  7.98 & 18.92 & 25.38 & 24.06 & 44.36 & 53.38 &  8.54 & 19.75 & 26.28 & 17.29 & 34.08 & 43.10 \\
SigLIP2 Giant          &  7.64 & 17.86 & 23.78 & 16.04 & 35.33 & 47.36 &  6.10 & 14.70 & 19.85 & 13.28 & 26.06 & 32.33 \\
MetaCLIP-2 Giant       & 17.17 & 34.88 & 43.53 & 36.59 & 57.39 & 65.91 & 14.21 & 30.57 & 39.43 & 34.58 & 55.63 & 67.41 \\
VideoCLIP-XL-v2        & 47.28 & 70.88 & 78.52 & 56.39 & 75.93 & 80.95 & 41.67 & 66.37 & 74.92 & 48.62 & 71.17 & 80.45 \\
Qwen3-VL-Emb-8B       & \textbf{80.27} & \textbf{93.99} & \textbf{96.75} & \textbf{98.49} & \textbf{99.74} & \textbf{99.74} & \textbf{77.61} & \textbf{92.94} & \textbf{96.14} & \textbf{95.48} & \textbf{100.0} & \textbf{100.0} \\
\midrule
\rowcolor{audiobg} \multicolumn{13}{c}{\textbf{Audio}} \\
MS-CLAP (2022)         &  0.13 &  0.54 &  0.90 &  2.00 &  5.76 &  9.02 &  0.13 &  0.48 &  0.81 &  1.75 &  5.01 &  8.02 \\
MS-CLAP (2023)         &  0.28 &  1.00 &  1.63 &  3.75 & 10.27 & 14.03 &  0.38 &  1.23 &  2.01 &  5.51 & 10.27 & 15.78 \\
LAION-CLAP             &  0.24 &  0.87 &  1.43 &  2.50 &  8.02 & 12.28 &  0.27 &  0.93 &  1.51 &  2.50 &  8.27 & 12.03 \\
M2D-CLAP               &  0.56 &  1.79 &  2.78 & 10.27 & 24.06 & 33.33 &  0.77 &  2.50 &  3.91 & 12.03 & 28.07 & 37.59 \\
GLAP                   &  0.53 &  1.60 &  2.46 &  5.76 & 15.03 & 23.55 &  0.72 &  1.98 &  2.99 &  8.27 & 24.31 & 32.33 \\
Aurola-7B              & \textbf{73.02} & \textbf{83.10} & \textbf{86.37} & \textbf{89.72} & \textbf{98.24} & \textbf{100.0} & \textbf{74.15} & \textbf{84.59} & \textbf{87.87} & \textbf{83.95} & \textbf{97.99} & \textbf{99.74} \\
\midrule
\rowcolor{unifiedbg} \multicolumn{13}{c}{\textbf{Vision{+}Audio}} \\
ImageBind              &  7.64 & 18.66 & 25.21 & 35.33 & 61.65 & 71.42 &  6.32 & 16.54 & 23.35 & 30.32 & 54.38 & 65.41 \\
LanguageBind           &  2.70 &  7.15 & 10.23 & 23.80 & 48.37 & 58.14 &  0.83 &  2.76 &  4.42 & 14.78 & 31.57 & 39.09 \\
Perception AV Large    & 26.48 & 42.32 & 49.05 & 49.12 & 74.68 & 81.95 & 26.06 & 46.37 & 55.13 & 40.10 & 66.91 & 75.18 \\
Wave-7B                & \textbf{65.51} & \textbf{83.50} & \textbf{88.26} & \textbf{91.23} & \textbf{99.75} & \textbf{100.0} & \textbf{66.22} & \textbf{83.99} & \textbf{88.57} & \textbf{93.73} & \textbf{99.50} & \textbf{100.0} \\
\bottomrule
\end{tabular}
}
}
\end{table}

\begin{table}[!t]
\centering
\caption{Query-based retrieval results (left) and baseline models (right). Results are clip-level Recall@$K$ (\%); \textbf{bold} indicates best within each group, and $^{\dagger}$ marks upgraded variants.}
\label{tab:baselines}
\label{tab:query_results}
{%
\scriptsize
\renewcommand{\arraystretch}{0.9}
\setlength{\tabcolsep}{2.6pt}
\resizebox{\textwidth}{!}{%
\begin{tabular}{l ccc ccc !{\vrule width 0.6pt} cc l}
\toprule
\multicolumn{7}{c!{\vrule width 0.6pt}}{\textbf{Query-based retrieval (Recall@$K$, \%)}} & \multicolumn{3}{c}{\textbf{Baseline models}} \\
\cmidrule(lr){1-7} \cmidrule(lr){8-10}
 & \multicolumn{3}{c}{Text$\to$Clip} & \multicolumn{3}{c}{Clip$\to$Text} & & & \\
\cmidrule(lr){2-4} \cmidrule(lr){5-7}
\textbf{Model} & R@1 & R@5 & R@10 & R@1 & R@5 & R@10 & \textbf{Type} & \textbf{Scale} & \textbf{Release} \\
\midrule
\rowcolor{visionbg} \multicolumn{10}{c}{\textbf{Vision}} \\
CLIP ViT-B/32              & 13.89 & 29.01 & 36.82 & 12.33 & 25.89 & 33.12 & Contrastive & 151M & ICML'21 \\
SigLIP2 Giant              & 33.98 & 56.92 & 65.17 & 22.90 & 43.59 & 52.75 & Contrastive & 1.1B & Google DeepMind'25.02 \\
MetaCLIP-2 Giant           & 33.09 & 57.18 & 66.20 & 21.46 & 41.51 & 51.06 & Contrastive & 2.5B & NeurIPS'25 \\
VideoCLIP-XL-v2$^{\dagger}$ & 29.57 & 53.77 & 63.60 & 31.53 & 55.47 & 65.21 & Contrastive & 1.4B & EMNLP'24 \\
Qwen3-VL-Emb-8B            & \textbf{60.82} & \textbf{84.80} & \textbf{90.41} & \textbf{56.80} & \textbf{81.17} & \textbf{88.07} & LLM-based   & 8B   & Tongyi Qwen'26.01 \\
\midrule
\rowcolor{audiobg} \multicolumn{10}{c}{\textbf{Audio}} \\
MS-CLAP (2022)             &  0.10 &  0.38 &  0.61 &  0.16 &  0.49 &  0.80 & Contrastive & 151M & ICASSP'23 \\
MS-CLAP (2023)$^{\dagger}$ &  0.30 &  0.92 &  1.45 &  0.31 &  1.03 &  1.57 & Contrastive & 151M & ICASSP'23 \\
LAION-CLAP                 &  0.19 &  0.62 &  1.01 &  0.18 &  0.60 &  0.97 & Contrastive & 151M & ICASSP'23 \\
M2D-CLAP                   &  0.44 &  1.37 &  2.10 &  0.55 &  1.71 &  2.62 & Contrastive & 300M & Interspeech'24 \\
GLAP                       &  0.63 &  1.56 &  2.25 &  0.59 &  1.43 &  2.05 & Contrastive & 300M & ICASSP'26 \\
Aurola-7B                  & \textbf{33.31} & \textbf{51.40} & \textbf{58.54} & \textbf{34.99} & \textbf{53.48} & \textbf{61.01} & LLM-based   & 7B   & Oxford VGG'26.02 \\
\midrule
\rowcolor{unifiedbg} \multicolumn{10}{c}{\textbf{Vision{+}Audio}} \\
ImageBind                  &  6.35 & 16.59 & 23.09 &  7.07 & 18.79 & 26.28 & Contrastive & 1.2B & CVPR'23 \\
LanguageBind               &  3.32 &  8.98 & 12.72 &  3.29 &  9.39 & 13.95 & Contrastive & 1.4B & ICLR'24 \\
Perception AV Large        &  7.79 & 18.51 & 24.87 &  9.37 & 21.93 & 29.01 & Contrastive & 2.5B & Meta AI'25.12 \\
Wave-7B                    & \textbf{42.63} & \textbf{67.63} & \textbf{76.26} & \textbf{47.69} & \textbf{71.66} & \textbf{79.78} & LLM-based   & 7B   & ICLR'26 \\
\bottomrule
\end{tabular}
}
}
\end{table}

\subsection{Query-Based Retrieval Results}
\label{sec:query_results}

Table~\ref{tab:query_results} presents query-based retrieval results using curated user-style queries at clip level. Compared with caption-based retrieval, several weaker contrastive baselines improve with shorter queries, likely because these inputs better match their limited text context and short-text alignment ability: CLIP ViT-B/32 rises from 7.98\% to 13.89\% T$\to$C R@1, MetaCLIP-2 from 17.17\% to 33.09\%, and SigLIP2 Giant from 7.64\% to 33.98\%. These gains, however, still do not close the gap to the strongest models. Conversely, caption-strong models drop under user-style queries, including Qwen3-VL-Emb-8B (80.27\%$\to$60.82\%), Aurola-7B (73.02\%$\to$33.31\%), and Wave-7B (65.51\%$\to$42.63\%). This shows that lower-information queries require finer-grained alignment between partial user intent and media content, making query-based retrieval a necessary complement to caption-based evaluation.

\begin{figure}[t]
\centering
\includegraphics[width=\textwidth]{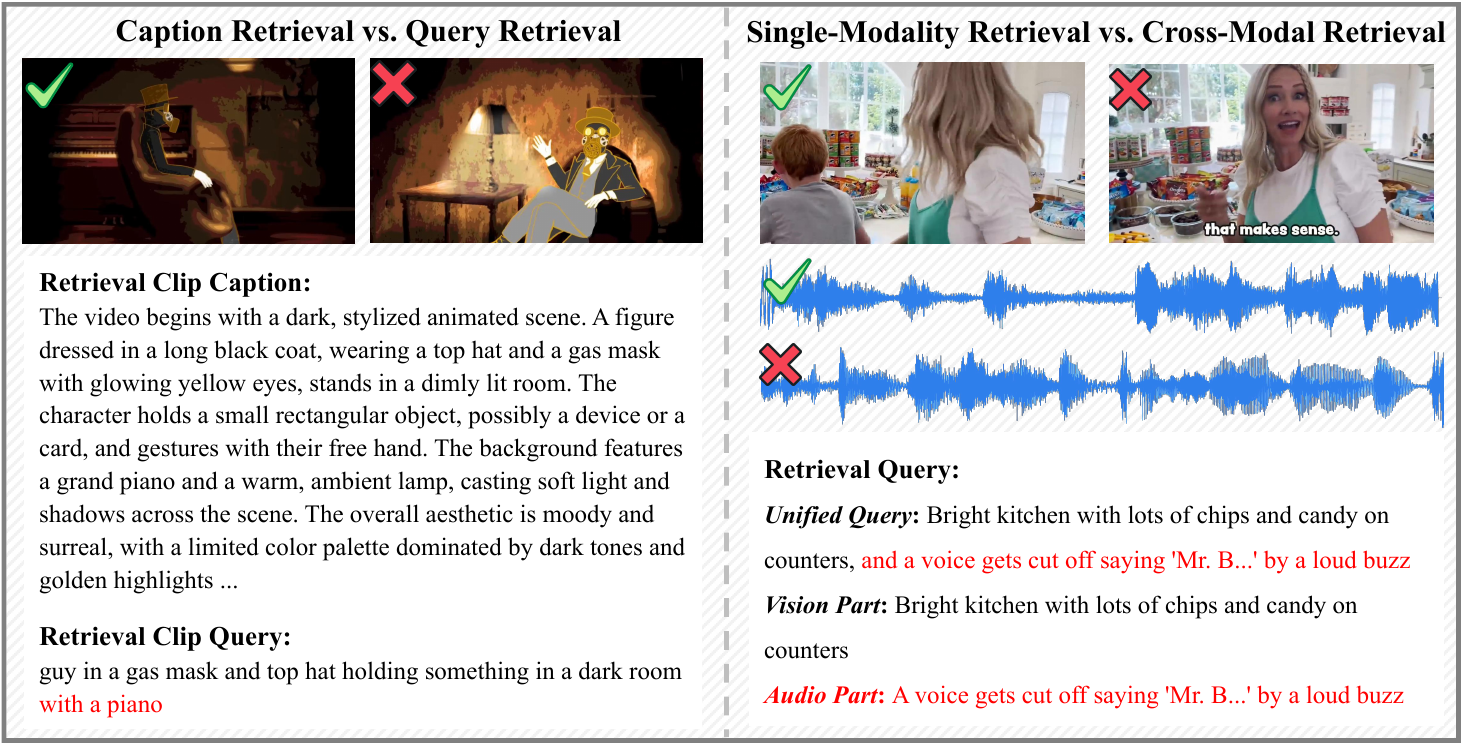}
\caption{Representative error cases. Left: a model retrieves the correct clip with a detailed caption but fails with a shorter user-style query. Right: a unified query requires matching both the visual scene and the audio event; the model retrieves a visually similar clip but ignores the mismatched audio cue.}
\label{fig:error_case}
\end{figure}

Appendix~\ref{app:modality_ablation} further probes unified audiovisual retrieval through modality ablations on both the media and text sides. The results reveal that the limited audio--language capability of current unified baselines is a major bottleneck: audio-only media and audio-only queries consistently lag behind their visual counterparts, and naive or insufficient fusion can even weaken retrieval. This suggests that future audiovisual retrieval models need not only stronger audio representations, but also better mechanisms for balancing and integrating visual and auditory features. These ablations also highlight the value of FLARE and the quality of its unified query construction: by jointly evaluating single-modal and fused audiovisual retrieval, the benchmark exposes fusion-level failures and cross-modal gains that unimodal evaluations cannot capture. Figure~\ref{fig:error_case} shows two representative retrieval errors, one from caption-vs-query retrieval and the other from unified audiovisual retrieval.

\section{Conclusion}
\label{sec:conclusion}

We introduced FLARE, a full-modality long-video audiovisual retrieval benchmark designed to better reflect realistic multimedia search. By combining long-form videos, fine-grained clip annotations, user-simulated queries, and a hard bimodal constraint for cross-modal queries, FLARE evaluates retrieval beyond short clips, single modalities, and caption-only settings. Our experiments across 15 representative models show that user-style queries substantially change model behavior, that LLM-based retrievers currently dominate caption-based evaluation, and that audio--language alignment remains a major bottleneck for unified audiovisual retrieval. We hope FLARE provides a focused testbed for developing retrieval models that can robustly align visual and auditory evidence with natural user intent.

\clearpage
\newpage
{
\small
\bibliographystyle{plainnat}
\bibliography{references}
}


\appendix


\section{Human Annotation Protocol}
\label{app:human_annotation}

Human review is integrated at multiple stages of the FLARE construction pipeline. This section details the annotator recruitment, task-specific protocols, inter-annotator agreement, and final quality evaluation.

\subsection{Annotator Recruitment and Training}
\label{app:annotators}

All human annotation tasks are carried out by the paper authors and recruited undergraduate students majoring in computer science. Before beginning any annotation task, each recruited annotator undergoes a training session in which the goals of the project, the annotation guidelines, and representative examples are explained. Recruited annotators are compensated at a rate of \$10 per hour, which meets or exceeds the local minimum wage standard. Throughout the annotation process, all disagreements that cannot be resolved between annotators are escalated to the paper authors for final adjudication.

\subsection{Clip Segmentation Verification}
\label{app:clip_verification}

After automated segmentation (Section~\ref{sec:clip_extraction}), clips still exceeding 2 minutes are routed to human review. Each such clip is independently reviewed by two annotators following a two-step protocol:

\begin{enumerate}[leftmargin=*, itemsep=2pt]
    \item \textbf{Video type classification.} The annotator first determines whether the clip is primarily \emph{visually driven} (e.g., sports, scenic transitions) or \emph{auditorily driven} (e.g., lectures, narrated content).
    \item \textbf{Segmentation decision.} Based on the identified type, the annotator segments the clip into coherent sub-clips according to the dominant modality's transition points. If neither visual nor auditory variation is sufficient to justify further splitting, the annotator flags the clip and reports it to the authors without forcing a split.
\end{enumerate}

After independent annotation, the two annotators discuss their segmentation decisions to produce a consensus result. Clips on which consensus cannot be reached are escalated to the authors, who make the final segmentation decision and filter out clips with insufficient audiovisual variation.

\subsection{Modality Triage Verification}
\label{app:modality_verification}

As described in Section~\ref{sec:clip_extraction}, clips exceeding one minute are scored by an LLM for audio importance ($s_{\text{aud}} \in [0, 10]$). To verify borderline cases, we select all clips classified as audio-dominant by the LLM but with $s_{\text{aud}} < 7$ and route them to human review. Two annotators independently label each clip as either \emph{visually dominant} or \emph{auditorily dominant}. Inter-annotator agreement on this binary classification task reaches Cohen's $\kappa = 0.75$ (substantial agreement). Clips with disagreeing labels are escalated to the authors for final decision.

\subsection{Caption Quality Assurance}
\label{app:caption_qa}

Caption quality assurance proceeds in two phases: threshold calibration and caption review.

\paragraph{Phase 1: Quality threshold calibration.}
To determine appropriate filtering thresholds for EVQAScore and BRACEScore, we randomly sample 500 captions per modality (vision, audio, unified). Each caption is independently reviewed by two annotators, who evaluate the following criteria:

\begin{itemize}[leftmargin=*, itemsep=2pt]
    \item Whether the caption contains content inconsistent with the video;
    \item Whether the caption omits key information present in the video;
    \item Whether the caption contains nonsensical or degenerate output.
\end{itemize}

A caption is marked as \emph{defective} if any of these issues (or other quality problems observed during review) is present. Inter-annotator agreement reaches Cohen's $\kappa = 0.68$ (substantial agreement). Disagreements are first discussed between annotators; unresolved cases are escalated to the authors. Analysis of the labeled data reveals that defective captions cluster almost entirely below EVQAScore $< 0.2$ for visual captions and BRACEScore $< 0.1$ for audio captions, which we adopt as the filtering thresholds $\theta_{\text{qa}}$.

\paragraph{Phase 2: Flagged caption review and correction.}
All captions falling below the quality thresholds (EVQAScore $< 0.2$ or BRACEScore $< 0.1$) or flagged by the LLM judge are subjected to human review. Each flagged caption is independently assessed by two annotators using the same criteria as Phase~1. Inter-annotator agreement in this phase reaches Cohen's $\kappa = 0.70$. Disagreements follow the same resolution protocol. For each caption confirmed as defective, the two annotators collaboratively revise the problematic content to produce a corrected caption.

\paragraph{Video-level caption review.}
Video-level captions produced by hierarchical merging (Section~\ref{sec:caption_merging}) are reviewed with additional criteria:

\begin{itemize}[leftmargin=*, itemsep=2pt]
    \item Whether any clip-level information is missing or contains nonsensical output;
    \item Whether transitions between consecutive clips are fluent and semantically coherent;
    \item Whether shared entities across clips are named consistently without introducing ambiguity.
\end{itemize}

Captions exhibiting any of these issues are revised or re-annotated by the reviewers.

\subsection{Final Quality Evaluation}
\label{app:final_quality}

To assess the overall quality of the finalized dataset, we conduct independent human evaluations on both captions and queries. Importantly, all samples used for final evaluation are drawn exclusively from captions that were \emph{not} manually revised, ensuring that the evaluation reflects the quality of the automated pipeline output.

\paragraph{Caption quality evaluation.}
We randomly sample 500 clip-level captions and 30 video-level captions from each modality (vision, audio, unified). Two annotators independently rate each caption on a 6-point Likert scale (0--5):

\begin{itemize}[leftmargin=*, itemsep=0pt]
    \item[\textbf{5}] -- Content fully matches the video; logically clear, fluent, with no errors or redundancy.
    \item[\textbf{4}] -- Content highly consistent with the video; minimal detail deviations that do not affect comprehension.
    \item[\textbf{3}] -- Content generally consistent but with noticeable omissions, minor errors, or unclear phrasing.
    \item[\textbf{2}] -- Partial inconsistency with the video; multiple errors or omissions that impair understanding.
    \item[\textbf{1}] -- Largely mismatched with the video; numerous errors or irrelevant content; poor overall quality.
    \item[\textbf{0}] -- Severely inconsistent or entirely unrelated to the video content.
\end{itemize}

The final mean score across all evaluated captions is \textbf{4.5}, with inter-annotator Cohen's $\kappa = 0.73$ (substantial agreement), confirming the high quality of the generated captions.

\paragraph{Query quality evaluation.}
We randomly sample 1,000 queries from each modality type (vision-only, audio-only, cross-modal). Two annotators independently rate each query on the same 6-point Likert scale (0--5), evaluating three dimensions: \emph{naturalness} (whether the query reflects realistic user search behavior), \emph{semantic correctness and clarity} (whether the query is accurate and unambiguous), and, for cross-modal queries, \emph{cross-modal completeness} (whether both visual and auditory information are adequately represented).

\begin{itemize}[leftmargin=*, itemsep=0pt]
    \item[\textbf{5}] -- Fully natural user-style query; clear, fluent, and free of errors or redundancy. For cross-modal queries: seamlessly integrates both visual and auditory information.
    \item[\textbf{4}] -- Generally natural with minor imprecisions or slight redundancy. For cross-modal queries: integrates both modalities well with minor gaps in fusion.
    \item[\textbf{3}] -- Somewhat natural but with noticeable unnaturalness, incomplete information, or minor logical issues. For cross-modal queries: covers both modalities but with loose integration or underrepresentation of one modality.
    \item[\textbf{2}] -- Clearly unnatural expression with significant clarity issues or information gaps. For cross-modal queries: poor modality integration with strong bias toward a single modality.
    \item[\textbf{1}] -- Does not resemble realistic user queries; confused or error-prone expression. For cross-modal queries: fails to effectively combine both modalities.
    \item[\textbf{0}] -- Incomprehensible or entirely unrelated to the target content. For cross-modal queries: no cross-modal information present.
\end{itemize}

The final mean score across all evaluated queries is \textbf{4.6}, with inter-annotator Cohen's $\kappa = 0.72$ (substantial agreement), confirming the quality and naturalness of the curated queries.


\section{Modality Ablation for Vision{+}Audio Models}
\label{app:modality_ablation}

The main evaluation (Table~\ref{tab:caption_results}) evaluates Vision{+}Audio models using unified audiovisual captions and media that jointly encode both modalities. A natural follow-up question is: \emph{how much does each individual modality contribute to the unified retrieval performance?} We design two complementary modality-ablation experiments to disentangle the roles of vision and audio within V{+}A models.

\subsection{Experimental Setup}
\label{app:modality_ablation_setup}

We evaluate all four V{+}A baselines---ImageBind, LanguageBind, Perception AV Large, and Wave-7B---under two ablation configurations:

\begin{itemize}[leftmargin=*, itemsep=2pt]
    \item \textbf{Ablation~A (Media-side).} We encode only a single modality (vision or audio) from the media side while keeping the unified text (both caption-based and query-based regimes) on the text side. This isolates each modality's contribution to the media embedding across all four retrieval directions (T$\to$C, C$\to$T, T$\to$V, V$\to$T).
    \item \textbf{Ablation~B (Text-side).} We use modality-specific queries (audio-only or vision-only) on the text side while keeping the full unified media encoding on the gallery side. Since single-modal queries are only defined per clip, this ablation is restricted to the query regime and the clip-level directions (T$\to$C, C$\to$T).
\end{itemize}

All other settings---similarity metric, candidate pools, and evaluation protocol---remain identical to Section~\ref{sec:exp_setup}. For the ``full unified media'' configuration, we note that ImageBind and LanguageBind provide no official recipe for combining per-modality embeddings into a single retrieval vector; following the standard late-fusion practice, we obtain the unified media embedding by average pooling of the $\ell_2$-normalized vision and audio embeddings. Perception AV Large and Wave-7B natively output a joint audiovisual embedding and are used as-is. Because media-side ablation involves a large number of configurations, we report Ablation~A in two complementary tables: Table~\ref{tab:ablation_a_r1} gives an R@1 overview comparing single-modal media against the full unified baseline, while Table~\ref{tab:ablation_a_full} reports the full R@1/5/10 results for the two single-modal configurations.

\subsection{Ablation A: Media-Side Single-Modal Retrieval}
\label{app:ablation_a}

\begin{table}[t]
\centering
\caption{Ablation~A (R@1 overview): Recall@1 (\%) of V{+}A models under full unified media versus single-modal media. Columns cover all four retrieval directions; text is the unified caption (Cap.) or unified query (Qry.). Since video-level evaluation depends only on the media side, its values are identical across caption and query regimes and are reported as a single column. Media types are colored for clarity: \colorbox{unifiedbg}{Full (V{+}A)}, \colorbox{visionbg}{Vision-only}, \colorbox{audiobg}{Audio-only}. \textbf{Bold} indicates best per column within each model group.}
\label{tab:ablation_a_r1}
\small
\setlength{\tabcolsep}{4pt}
\begin{tabular}{l l cc cc c c}
\toprule
 & & \multicolumn{2}{c}{T$\to$Clip} & \multicolumn{2}{c}{Clip$\to$Text} & T$\to$Video & Video$\to$Text \\
\cmidrule(lr){3-4} \cmidrule(lr){5-6} \cmidrule(lr){7-7} \cmidrule(lr){8-8}
Model & Media & Cap. & Qry. & Cap. & Qry. & R@1 & R@1 \\
\midrule
\multirow{3}{*}{ImageBind}
  & \cellcolor{unifiedbg} Full (V{+}A) &  7.64 &  6.35 &  6.32 &  7.07 & 35.33 & 30.32 \\
  & \cellcolor{visionbg}  Vision-only  & \textbf{19.04} & \textbf{12.87} & \textbf{19.06} & \textbf{15.50} & \textbf{52.13} & \textbf{51.38} \\
  & \cellcolor{audiobg}   Audio-only   &  0.17 &  0.31 &  0.14 &  0.45 &  7.02 &  5.76 \\
\midrule
\multirow{3}{*}{LanguageBind}
  & \cellcolor{unifiedbg} Full (V{+}A) &  2.70 &  3.32 &  0.83 &  3.29 & 23.80 & 14.78 \\
  & \cellcolor{visionbg}  Vision-only  & \textbf{19.94} & \textbf{15.52} & \textbf{20.00} & \textbf{20.22} & \textbf{49.12} & \textbf{48.37} \\
  & \cellcolor{audiobg}   Audio-only   &  0.05 &  0.14 &  0.04 &  0.11 &  1.50 &  1.00 \\
\midrule
\multirow{3}{*}{Perception AV}
  & \cellcolor{unifiedbg} Full (V{+}A) & \textbf{26.48} & \textbf{7.79} & 26.06 &  9.37 & 49.12 & 40.10 \\
  & \cellcolor{visionbg}  Vision-only  & 24.55 &  7.22 & \textbf{26.75} & \textbf{10.69} & \textbf{58.90} & \textbf{51.13} \\
  & \cellcolor{audiobg}   Audio-only   &  5.49 &  0.37 &  5.64 &  2.17 & 16.54 & 14.79 \\
\midrule
\multirow{3}{*}{Wave-7B}
  & \cellcolor{unifiedbg} Full (V{+}A) & \textbf{65.51} & \textbf{42.63} & \textbf{66.22} & \textbf{47.69} & \textbf{91.23} & \textbf{93.73} \\
  & \cellcolor{visionbg}  Vision-only  & 27.84 & 16.80 & 32.45 & 20.73 & \textbf{91.23} & 88.72 \\
  & \cellcolor{audiobg}   Audio-only   & 13.27 &  7.58 & 15.11 &  8.07 & 64.16 & 50.88 \\
\bottomrule
\end{tabular}
\end{table}

Table~\ref{tab:ablation_a_r1} reveals a striking contrast between per-modality embedding quality and fused retrieval performance. For ImageBind and LanguageBind, \emph{no official recipe is provided for combining vision and audio media into a single retrieval embedding}; following common practice, we therefore fuse the two modality-specific embeddings by average pooling. Under this fusion, vision-only media dramatically \emph{outperforms} the fused representation across all six directions: LanguageBind's vision-only T$\to$C R@1 (19.94\%) is 7.4$\times$ its fused result (2.70\%), and ImageBind shows the same 2.5$\times$ gap (19.04\% vs.\ 7.64\%). The root cause is visible in the per-modality columns: the gulf between vision-only and audio-only embedding quality (e.g., LanguageBind 19.94\% vs.\ 0.05\% on T$\to$C R@1) is so large that simple pooling is dominated by the weaker audio component rather than being complemented by it. In other words, although these models \emph{produce} embeddings for multiple modalities, the severe imbalance between their per-modality embedding qualities causes naive fusion to actively degrade retrieval. Perception AV Large, trained with a designed audiovisual objective, exhibits a much smaller imbalance and thus a fused representation that is close to its vision-only performance; only Wave-7B, benefiting from LLM-scale joint training, yields a fused embedding that clearly surpasses either single-modal variant at clip level (e.g., 65.51\% fused vs.\ 27.84\% vision-only on T$\to$C R@1). This pattern highlights a core value of FLARE: a retrieval benchmark that jointly evaluates single-modal and fused audiovisual retrieval is necessary to expose fusion-level failures that per-modality evaluations alone would never surface, and to quantify how well a V{+}A model truly integrates its two modalities---an angle that unimodal benchmarks fundamentally cannot capture.

\begin{table}[t]
\centering
\caption{Ablation~A (detailed R@1/5/10): Clip-level retrieval metrics of single-modal media against unified text, under both caption and query regimes. Video-level results are already summarized by R@1 in Table~\ref{tab:ablation_a_r1}. \textbf{Bold} indicates best per column within each media group.}
\label{tab:ablation_a_full}
\small
\setlength{\tabcolsep}{3pt}
\resizebox{\columnwidth}{!}{
\begin{tabular}{l l ccc ccc ccc ccc}
\toprule
 & & \multicolumn{6}{c}{Caption regime} & \multicolumn{6}{c}{Query regime} \\
\cmidrule(lr){3-8} \cmidrule(lr){9-14}
 & & \multicolumn{3}{c}{T$\to$Clip} & \multicolumn{3}{c}{Clip$\to$Text} & \multicolumn{3}{c}{T$\to$Clip} & \multicolumn{3}{c}{Clip$\to$Text} \\
\cmidrule(lr){3-5} \cmidrule(lr){6-8} \cmidrule(lr){9-11} \cmidrule(lr){12-14}
Model & Media & R@1 & R@5 & R@10 & R@1 & R@5 & R@10 & R@1 & R@5 & R@10 & R@1 & R@5 & R@10 \\
\midrule
\rowcolor{visionbg} \multicolumn{14}{c}{\textbf{Vision-Only Media}} \\
ImageBind      & V-only & 19.04 & 38.10 & 46.81 & 19.06 & 38.27 & 47.32 & 12.87 & 29.86 & 39.24 & 15.50 & 33.99 & 44.48 \\
LanguageBind   & V-only & 19.94 & 39.59 & 48.47 & 20.00 & 39.26 & 48.06 & 15.52 & 34.82 & 44.82 & 20.22 & 42.13 & 52.22 \\
Perception AV  & V-only & 24.55 & 42.23 & 49.71 & 26.75 & 50.66 & 60.97 &  7.22 & 18.54 & 25.21 & 10.69 & 25.71 & 34.49 \\
Wave-7B        & V-only & \textbf{27.84} & \textbf{51.37} & \textbf{61.57} & \textbf{32.45} & \textbf{55.98} & \textbf{65.47} & \textbf{16.80} & \textbf{37.54} & \textbf{47.92} & \textbf{20.73} & \textbf{42.29} & \textbf{53.06} \\
\midrule
\rowcolor{audiobg} \multicolumn{14}{c}{\textbf{Audio-Only Media}} \\
ImageBind      & A-only &  0.17 &  0.69 &  1.17 &  0.14 &  0.60 &  1.05 &  0.31 &  1.18 &  1.98 &  0.45 &  1.56 &  2.62 \\
LanguageBind   & A-only &  0.05 &  0.19 &  0.32 &  0.04 &  0.12 &  0.21 &  0.14 &  0.50 &  0.78 &  0.11 &  0.42 &  0.69 \\
Perception AV  & A-only &  5.49 & 13.17 & 18.01 &  5.64 & 13.20 & 17.96 &  0.37 &  1.33 &  2.28 &  2.17 &  6.28 &  9.59 \\
Wave-7B        & A-only & \textbf{13.27} & \textbf{27.18} & \textbf{34.35} & \textbf{15.11} & \textbf{30.99} & \textbf{39.06} &  \textbf{7.58} & \textbf{18.41} & \textbf{24.80} &  \textbf{8.07} & \textbf{19.58} & \textbf{26.39} \\
\bottomrule
\end{tabular}
}
\end{table}

Table~\ref{tab:ablation_a_full} reports the full R@1/5/10 breakdown for the two single-modal media configurations across both regimes. Several patterns become visible at higher $K$. First, the ranking among models on vision-only media is consistent with the main evaluation: Wave-7B $>$ Perception AV $>$ LanguageBind $\approx$ ImageBind. Second, the \textbf{caption$\to$query degradation} observed in the main evaluation persists and is often amplified under single-modal media: Perception AV drops from 24.55\% to 7.22\% on vision-only T$\to$C R@1, a 3.4$\times$ reduction comparable in magnitude to the full-unified regime. Third, for audio-only media, the two contrastive V{+}A baselines (ImageBind, LanguageBind) are essentially inoperative at clip level ($<$1\% R@10 under queries), while Perception AV and especially Wave-7B (13.27\% T$\to$C R@1, 34.35\% R@10 under captions) demonstrate that a non-trivial audio--language signal can be extracted from V{+}A media when trained at sufficient scale.

\subsection{Ablation B: Text-Side Single-Modal Queries}
\label{app:ablation_b}

\begin{table}[t]
\centering
\caption{Ablation~B: Clip-level retrieval with single-modal queries against full unified media (Recall@$K$, \%). Since single-modal queries are defined only per clip and video-level retrieval depends only on the media side, the video-level results coincide with the main evaluation and are omitted here. Query types are colored for clarity: \colorbox{unifiedbg}{Full unified}, \colorbox{visionbg}{Vision-only}, \colorbox{audiobg}{Audio-only}. \textbf{Bold} indicates best per column within each model group.}
\label{tab:ablation_b}
\small
\setlength{\tabcolsep}{3.5pt}
\begin{tabular}{l l ccc ccc}
\toprule
 & & \multicolumn{3}{c}{Text$\to$Clip} & \multicolumn{3}{c}{Clip$\to$Text} \\
\cmidrule(lr){3-5} \cmidrule(lr){6-8}
Model & Query & R@1 & R@5 & R@10 & R@1 & R@5 & R@10 \\
\midrule
\multirow{3}{*}{ImageBind}
  & \cellcolor{unifiedbg} Full unified & \textbf{6.35} & \textbf{16.59} & \textbf{23.09} & \textbf{7.07} & \textbf{18.79} & \textbf{26.28} \\
  & \cellcolor{visionbg}  Vision-only  & 4.20 & 12.24 & 17.85 & 4.08 & 11.63 & 17.17 \\
  & \cellcolor{audiobg}   Audio-only   & 0.41 &  1.23 &  1.90 & 0.78 &  2.19 &  3.47 \\
\midrule
\multirow{3}{*}{LanguageBind}
  & \cellcolor{unifiedbg} Full unified & \textbf{3.32} & \textbf{8.98} & \textbf{12.72} & \textbf{3.29} & \textbf{9.39} & \textbf{13.95} \\
  & \cellcolor{visionbg}  Vision-only  & 1.53 & 4.43 &  6.58 & 1.56 & 5.11 &  8.13 \\
  & \cellcolor{audiobg}   Audio-only   & 0.30 & 0.99 &  1.55 & 0.34 & 1.18 &  1.85 \\
\midrule
\multirow{3}{*}{Perception AV}
  & \cellcolor{unifiedbg} Full unified & \textbf{7.79} & \textbf{18.51} & \textbf{24.87} & \textbf{9.37} & \textbf{21.93} & \textbf{29.01} \\
  & \cellcolor{visionbg}  Vision-only  & 3.02 &  8.76 & 12.69 & 3.09 &  8.53 & 12.07 \\
  & \cellcolor{audiobg}   Audio-only   & 0.69 &  2.24 &  3.40 & 1.05 &  3.03 &  4.76 \\
\midrule
\multirow{3}{*}{Wave-7B}
  & \cellcolor{unifiedbg} Full unified & \textbf{42.63} & \textbf{67.63} & \textbf{76.26} & \textbf{47.69} & \textbf{71.66} & \textbf{79.78} \\
  & \cellcolor{visionbg}  Vision-only  & 13.76 & 32.07 & 41.91 & 20.74 & 42.42 & 52.96 \\
  & \cellcolor{audiobg}   Audio-only   &  6.41 & 14.62 & 19.50 &  9.08 & 20.01 & 26.01 \\
\bottomrule
\end{tabular}
\end{table}

Table~\ref{tab:ablation_b} measures the opposite ablation: media is kept as the full unified embedding, and the text query is restricted to a single modality. Across all four models, \textbf{both single-modal queries substantially underperform the full unified query}, and vision queries consistently outperform audio queries. For Wave-7B, vision-only queries recover only 32.3\% of the full unified query T$\to$C R@1 (13.76\% vs.\ 42.63\%), and audio-only queries recover merely 15.0\% (6.41\% vs.\ 42.63\%). For the contrastive models, the gap is even more pronounced: Perception AV's audio-only queries collapse to 0.69\% T$\to$C R@1 against a full unified reference of 7.79\%. The consistent ordering \emph{Full unified $>$ Vision-only $>$ Audio-only} indicates that (i) the unified-media embeddings cannot be reliably retrieved from audio-dominated language descriptions alone, exposing a vision-biased fusion in all four models, and (ii) vision and audio contribute \emph{non-redundant} textual cues---neither single-modal query can match joint retrieval performance, validating the need for cross-modal queries in FLARE.


\section{Prompt Templates}
\label{app:prompts}

\begin{figure}[p]
\centering
\includegraphics[width=\textwidth]{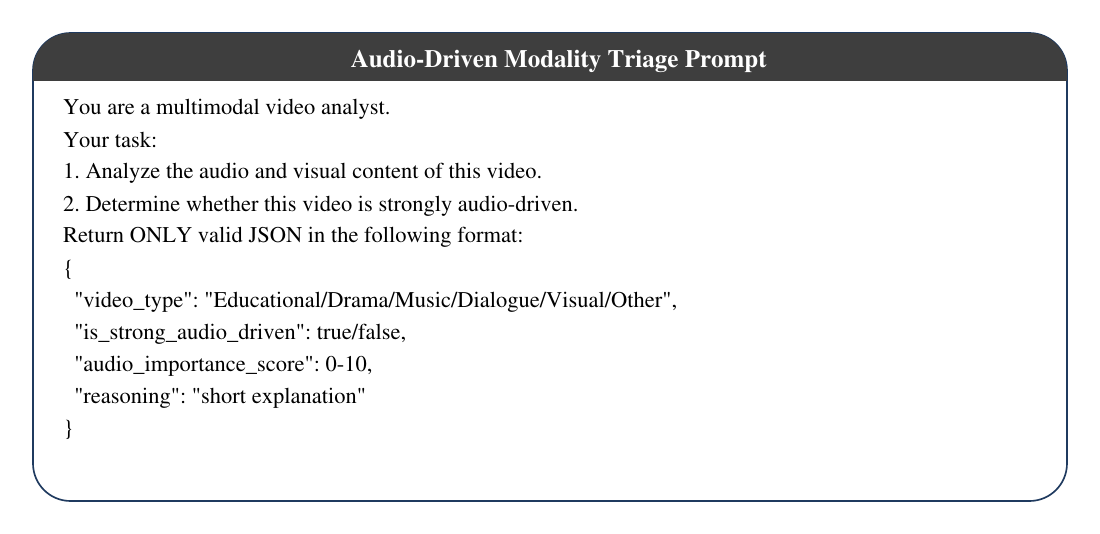}
\caption{Prompt for audio-driven modality triage.}
\label{fig:prompt_audio_triage}
\end{figure}

\begin{figure}[p]
\centering
\includegraphics[width=\textwidth]{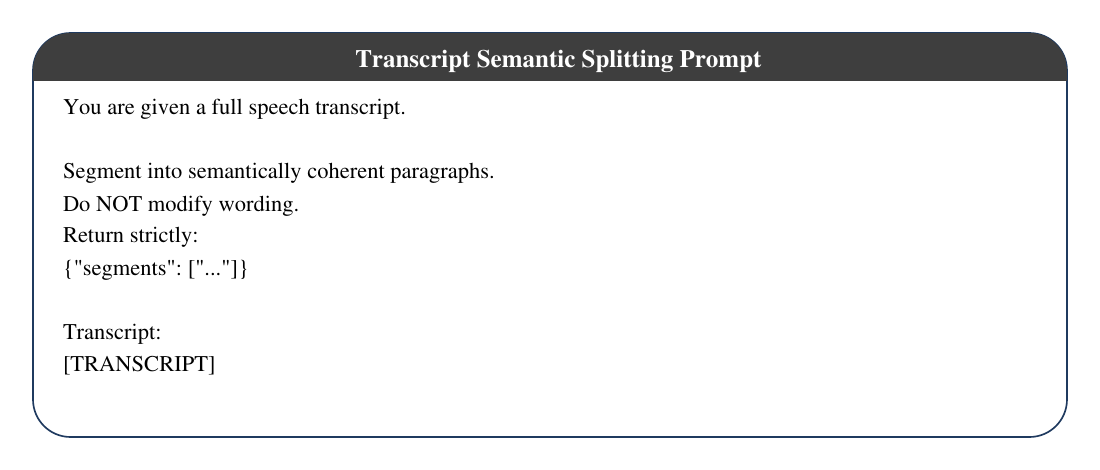}
\caption{Prompt for transcript-based semantic splitting.}
\label{fig:prompt_semantic_split}
\end{figure}

\begin{figure}[p]
\centering
\includegraphics[width=\textwidth]{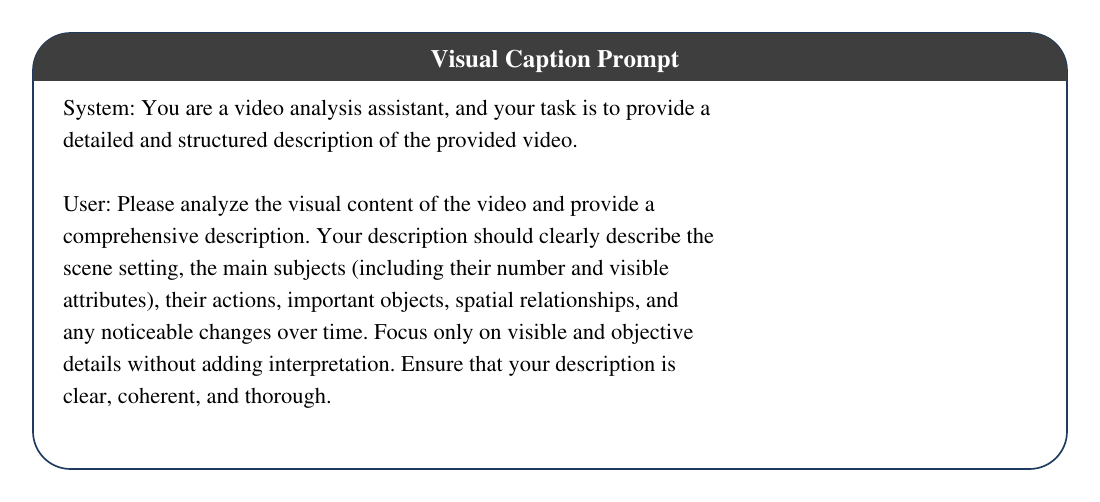}
\caption{Prompt for visual clip-level caption generation.}
\label{fig:prompt_visual_caption}
\end{figure}

\begin{figure}[p]
\centering
\includegraphics[width=\textwidth]{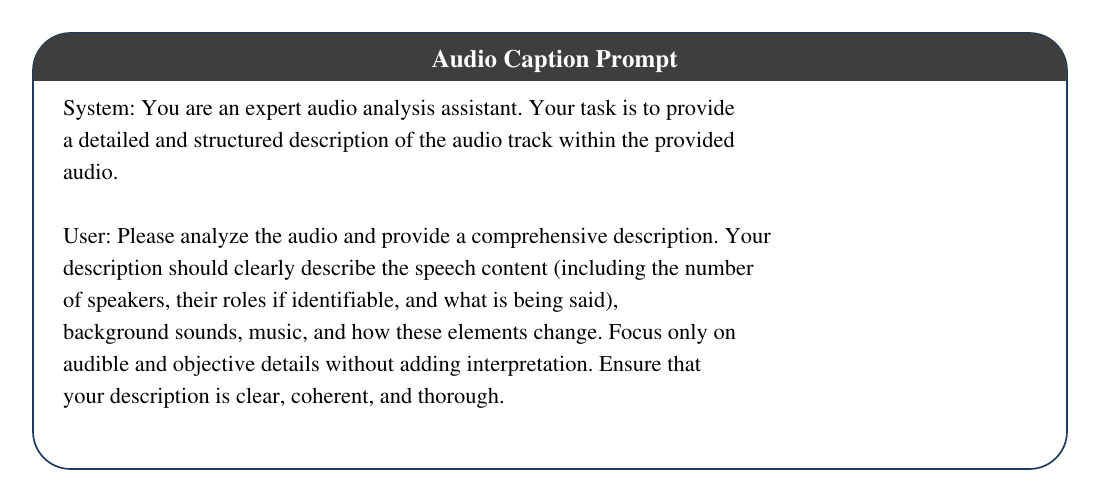}
\caption{Prompt for audio clip-level caption generation.}
\label{fig:prompt_audio_caption}
\end{figure}

\begin{figure}[p]
\centering
\includegraphics[width=\textwidth]{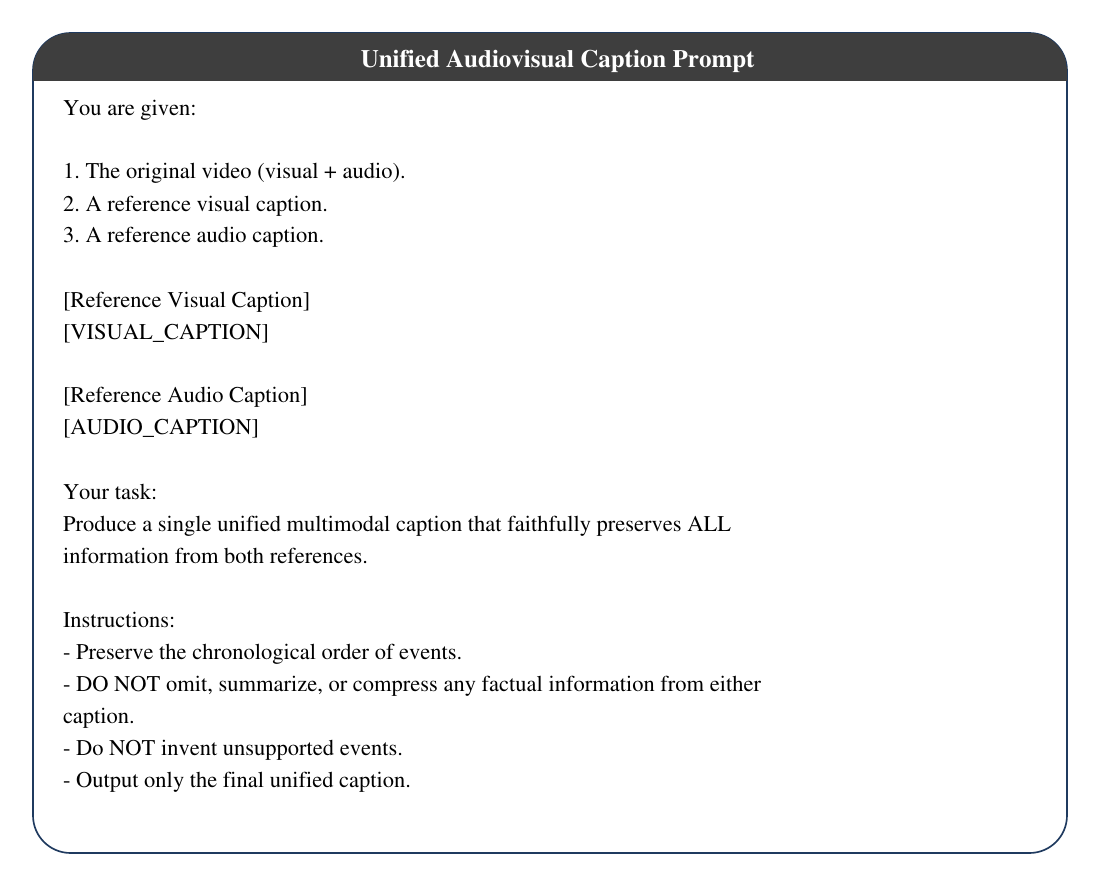}
\caption{Prompt for unified audiovisual caption generation.}
\label{fig:prompt_unified_caption}
\end{figure}

\begin{figure}[p]
\centering
\includegraphics[width=\textwidth,height=\textheight,keepaspectratio]{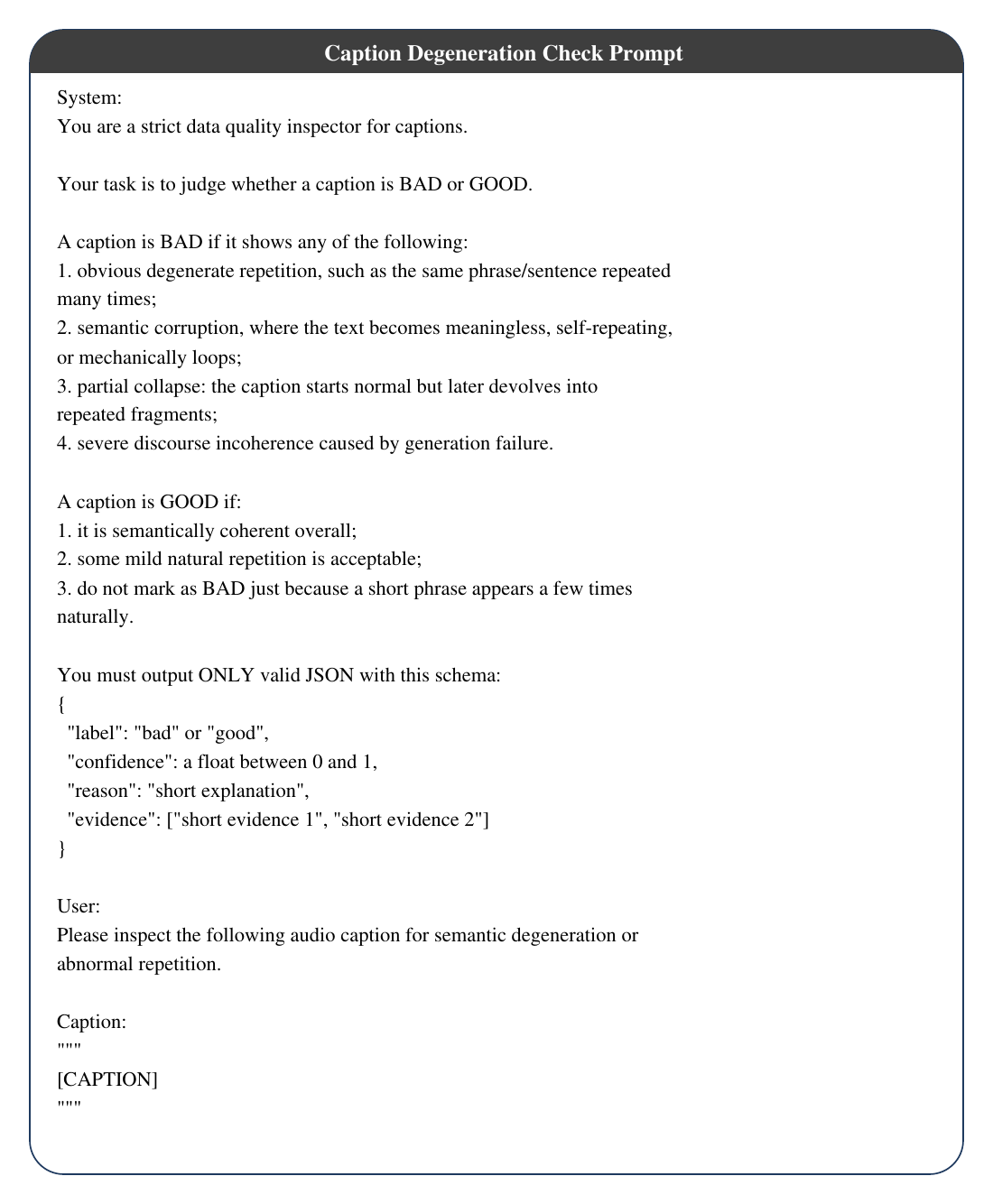}
\caption{Prompt for caption degeneration checking.}
\label{fig:prompt_caption_judge}
\end{figure}

\begin{figure}[p]
\centering
\includegraphics[width=\textwidth]{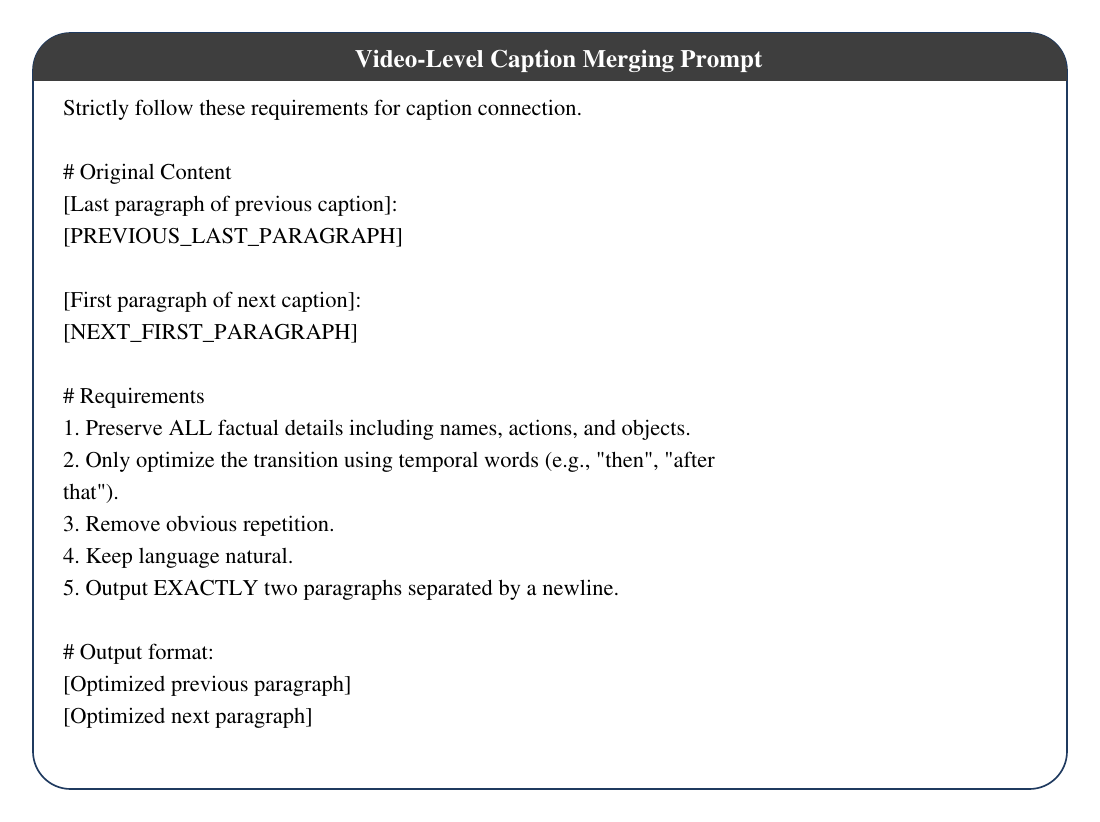}
\caption{Prompt for video-level caption merging.}
\label{fig:prompt_video_level_merge}
\end{figure}

\begin{figure}[p]
\centering
\includegraphics[width=\textwidth,height=\textheight,keepaspectratio]{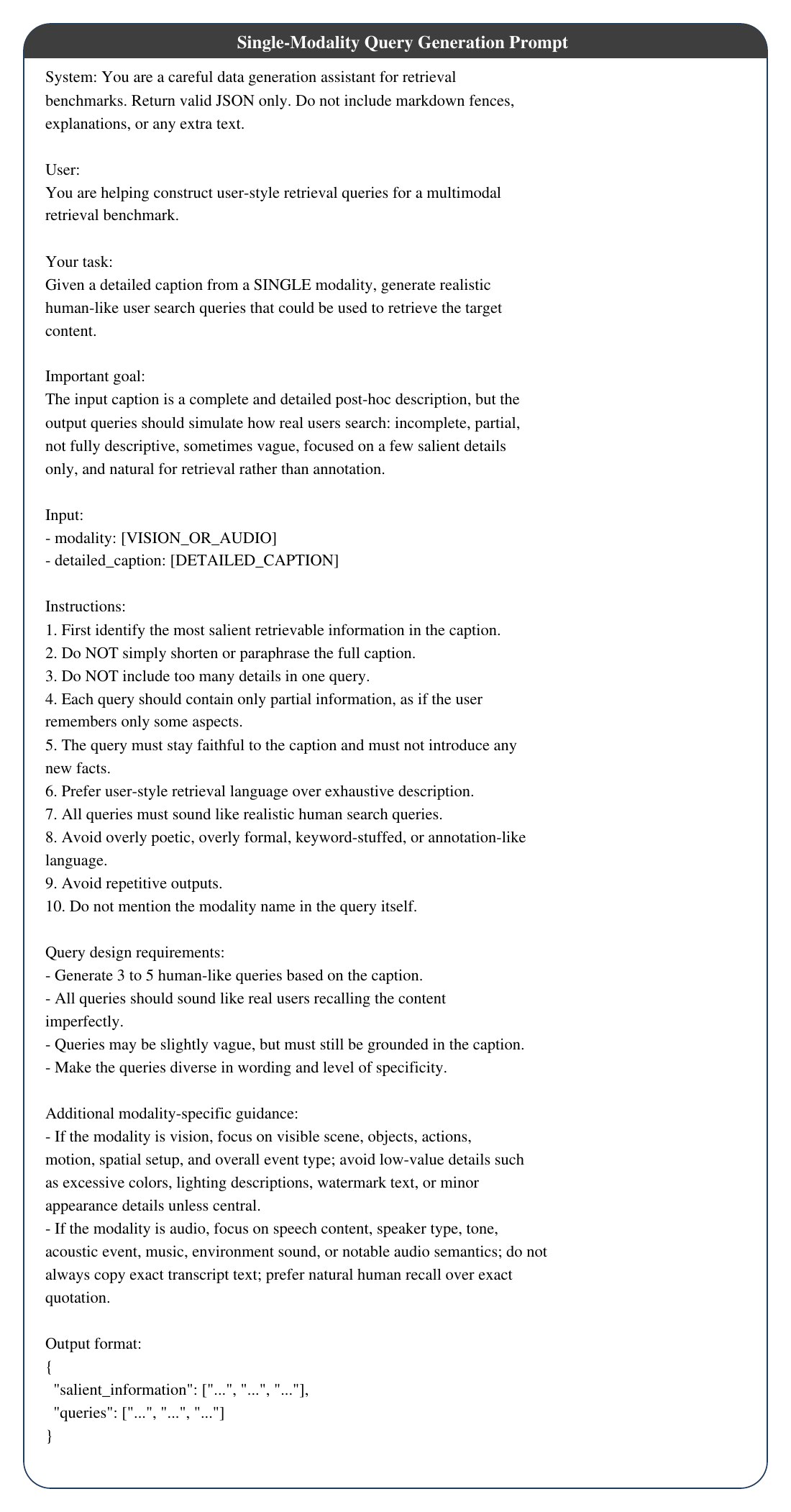}
\caption{Prompt for single-modality query generation.}
\label{fig:prompt_single_query}
\end{figure}

\begin{figure}[p]
\centering
\includegraphics[width=\textwidth,height=\textheight,keepaspectratio]{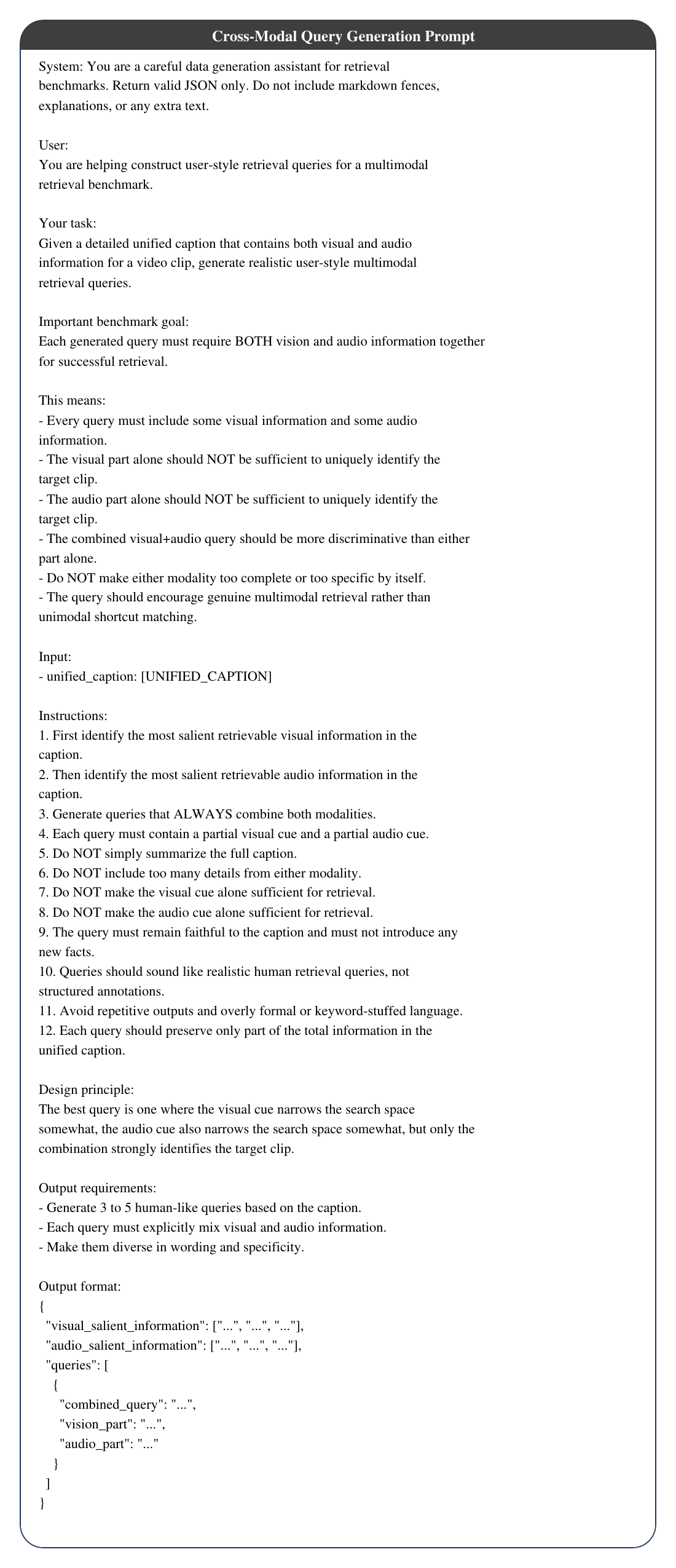}
\caption{Prompt for cross-modal query generation.}
\label{fig:prompt_cross_modal_query}
\end{figure}

\clearpage
\newpage

\section{Limitations}
\label{sec:limitations}

FLARE is built from curated Video-MME videos, which improves source quality but may not cover all domains, languages, cultures, or low-quality user-generated content. Its user-style queries are simulated from captions rather than collected from deployed search logs, and automated annotations may still contain residual omissions or biases despite quality checks and human review. In addition, several construction stages rely primarily on Qwen-series models, which may introduce model-specific preferences in captions, query phrasing, and filtering decisions. Finally, our evaluation focuses on retrieval accuracy with fixed public checkpoints, leaving efficiency, robustness, and deployment-time user satisfaction for future work.

\section{Ethics}
\label{sec:ethics}

FLARE uses existing video sources and public models under their applicable terms, and the release will document sources, derived annotations, licenses, and intended use. Human annotators performed only non-interventional annotation and quality assessment tasks, which were determined to be institutionally exempt from formal IRB review. No behavioral experiments, interventions, or collection of personal or sensitive data from annotators were involved.

\section{Broader Impact}
\label{sec:broader_impact}

FLARE provides a more realistic testbed for long-video audiovisual retrieval and may help bridge the gap between caption-based evaluation and real user-style search behavior. At the same time, advances in video retrieval systems could potentially be misused for invasive search, large-scale surveillance, or biased content discovery. To mitigate these risks, FLARE is intended solely for research and benchmark evaluation purposes. The dataset is constructed from existing video sources under their applicable licenses and terms of use, and no personally sensitive annotations are intentionally collected.

\clearpage
\newpage
\section*{NeurIPS Paper Checklist}

\begin{enumerate}

\item {\bf Claims}
    \item[] Question: Do the main claims made in the abstract and introduction accurately reflect the paper's contributions and scope?
    \item[] Answer: \answerYes{}.
    \item[] Justification: The abstract and Section~\ref{sec:introduction} state the benchmark scope, construction scale, evaluation regimes, and empirical findings. These claims are supported by the dataset construction details in Section~\ref{sec:dataset_construction} and the experimental results in Section~\ref{sec:experiments}.

\item {\bf Limitations}
    \item[] Question: Does the paper discuss the limitations of the work performed by the authors?
    \item[] Answer: \answerYes{}.
    \item[] Justification: The paper includes a dedicated limitations discussion in Section~\ref{sec:limitations}. It discusses the scope and assumptions of the benchmark, including the use of source videos, automated generation, human review, and evaluation coverage.

\item {\bf Theory assumptions and proofs}
    \item[] Question: For each theoretical result, does the paper provide the full set of assumptions and a complete (and correct) proof?
    \item[] Answer: \answerNA{}.
    \item[] Justification: The paper introduces a benchmark and empirical evaluation protocol rather than theoretical results. The equations in the paper define construction and filtering procedures and do not require formal proofs.

\item {\bf Experimental result reproducibility}
    \item[] Question: Does the paper fully disclose all the information needed to reproduce the main experimental results of the paper to the extent that it affects the main claims and/or conclusions of the paper (regardless of whether the code and data are provided or not)?
    \item[] Answer: \answerYes{}.
    \item[] Justification: Section~\ref{sec:dataset_construction} describes the benchmark construction pipeline, and Section~\ref{sec:exp_setup} specifies the retrieval protocol, metrics, directions, model set, and use of public checkpoints. The appendix further provides human annotation details and prompt templates needed to verify and reproduce the data-generation process.

\item {\bf Open access to data and code}
    \item[] Question: Does the paper provide open access to the data and code, with sufficient instructions to faithfully reproduce the main experimental results, as described in supplemental material?
    \item[] Answer: \answerYes{}.
    \item[] Justification: Our code and data are released at \url{https://anonymous.4open.science/r/FLARE-950E/} and \url{https://huggingface.co/datasets/AnonymousFLARE/FLARE}. The release will preserve anonymity during review and will be de-anonymized upon acceptance.

\item {\bf Experimental setting/details}
    \item[] Question: Does the paper specify all the training and test details (e.g., data splits, hyperparameters, how they were chosen, type of optimizer) necessary to understand the results?
    \item[] Answer: \answerYes{}.
    \item[] Justification: Section~\ref{sec:exp_setup} specifies the evaluation metric, similarity function, retrieval directions, caption/query regimes, candidate galleries, and baseline checkpoints. The work evaluates pretrained retrieval models without additional training, so optimizer and training hyperparameters are not applicable.

\item {\bf Experiment statistical significance}
    \item[] Question: Does the paper report error bars suitably and correctly defined or other appropriate information about the statistical significance of the experiments?
    \item[] Answer: \answerNo{}.
    \item[] Justification: The reported results are deterministic evaluations of fixed public checkpoints on a fixed benchmark gallery, and the paper reports Recall@K rather than repeated-run statistics. We therefore do not report error bars, while using multiple retrieval directions, regimes, and ablations to support the empirical conclusions.

\item {\bf Experiments compute resources}
    \item[] Question: For each experiment, does the paper provide sufficient information on the computer resources (type of compute workers, memory, time of execution) needed to reproduce the experiments?
    \item[] Answer: \answerYes{}.
    \item[] Justification: The human resources used in our annotation process are described in Section~\ref{app:human_annotation}. For both the dataset construction and evaluation stages, we used 8$\times$H20 GPUs, with a total computational cost of approximately 300 GPU hours.
    
\item {\bf Code of ethics}
    \item[] Question: Does the research conducted in the paper conform, in every respect, with the NeurIPS Code of Ethics \url{https://neurips.cc/public/EthicsGuidelines}?
    \item[] Answer: \answerYes{}.
    \item[] Justification: This research uses existing video sources and publicly available models in accordance with their respective licenses and terms of use, incorporates compensated human review during the annotation process, and is intended solely for benchmark construction and evaluation purposes. We adhere to the NeurIPS Code of Ethics; additional details are provided in Section~\ref{sec:ethics}.

\item {\bf Broader impacts}
    \item[] Question: Does the paper discuss both potential positive societal impacts and negative societal impacts of the work performed?
    \item[] Answer: \answerYes{}.
    \item[] Justification: We discuss the related broader impact considerations in Section~\ref{sec:broader_impact}.
    
\item {\bf Safeguards}
    \item[] Question: Does the paper describe safeguards that have been put in place for responsible release of data or models that have a high risk for misuse (e.g., pre-trained language models, image generators, or scraped datasets)?
    \item[] Answer: \answerYes{}.
    \item[] Justification: To support the responsible release of the benchmark, we follow the licenses and terms of use associated with the original video sources and publicly available models. FLARE is intended solely for research and benchmark evaluation purposes. We release only the annotations and metadata necessary for evaluation and do not intentionally collect personally sensitive information. 

\item {\bf Licenses for existing assets}
    \item[] Question: Are the creators or original owners of assets (e.g., code, data, models), used in the paper, properly credited and are the license and terms of use explicitly mentioned and properly respected?
    \item[] Answer: \answerYes{}.
    \item[] Justification: The paper cites the existing datasets, tools, and models used in construction and evaluation, including Video-MME, PySceneDetect, captioning models, scoring models, and retrieval baselines. Their licenses and terms of use will be documented and respected in the release materials.

\item {\bf New assets}
    \item[] Question: Are new assets introduced in the paper well documented and is the documentation provided alongside the assets?
    \item[] Answer: \answerYes{}.
    \item[] Justification: We have provided the related materials following the submission guidelines.

\item {\bf Crowdsourcing and research with human subjects}
    \item[] Question: For crowdsourcing experiments and research with human subjects, does the paper include the full text of instructions given to participants and screenshots, if applicable, as well as details about compensation (if any)? 
    \item[] Answer: \answerYes{}.
    \item[] Justification: Appendix~\ref{app:human_annotation} describes annotator recruitment, training, task protocols, compensation, agreement, and adjudication.

\item {\bf Institutional review board (IRB) approvals or equivalent for research with human subjects}
    \item[] Question: Does the paper describe potential risks incurred by study participants, whether such risks were disclosed to the subjects, and whether Institutional Review Board (IRB) approvals (or an equivalent approval/review based on the requirements of your country or institution) were obtained?
    \item[] Answer: \answerYes{}.
    \item[] Justification: Human annotators were limited to non-interventional annotation and quality assessment tasks, which were determined to be institutionally exempt from formal IRB review. The study did not involve behavioral experiments, interventions, or collection of personal or sensitive data from annotators.

\item {\bf Declaration of LLM usage}
    \item[] Question: Does the paper describe the usage of LLMs if it is an important, original, or non-standard component of the core methods in this research? Note that if the LLM is used only for writing, editing, or formatting purposes and does \emph{not} impact the core methodology, scientific rigor, or originality of the research, declaration is not required.
    \item[] Answer: \answerYes{}.
    \item[] Justification: LLMs and multimodal LLMs are core components of the benchmark construction pipeline, including modality triage, transcript segmentation, caption generation, caption merging, query generation, and quality checks. These uses are described in Section~\ref{sec:dataset_construction}, with prompt templates provided in Appendix~\ref{app:prompts}.

\end{enumerate}
\clearpage
\newpage


\end{document}